\documentclass[AMA,STIX1COL,doublespace]{WileyNJD-v2}

\usepackage{booktabs}
\usepackage{xcolor}
\usepackage{ulem}
\usepackage{pdflscape}

\articletype{Research article}

\providecommand{\tightlist}{%
  \setlength{\itemsep}{0pt}\setlength{\parskip}{0pt}}

\begin{document}

\title{Impact of model misspecification in shared frailty survival models}

\author[]{Alessandro Gasparini\textsuperscript{$1,\dagger$}}
\author[2]{Mark S. Clements}
\author[1]{Keith R. Abrams}
\author[1]{Michael J. Crowther}

\authormark{Gasparini \emph{et al}}

\address[1]{Biostatistics Research Group, Department of Health Sciences, University
of Leicester, Leicester, United Kingdom}
\address[2]{Department of Medical Epidemiology and Biostatistics, Karolinska
Institutet, Stockholm, Sweden}

\corres{\textsuperscript{$\dagger$} Correspondence to: Alessandro Gasparini,
Department of Health Sciences, University of Leicester, Centre for
Medicine, University Road, Leicester, LE1 7RH, United Kingdom. E-mail:
\href{mailto:ag475@leicester.ac.uk}{\nolinkurl{ag475@leicester.ac.uk}}}

\abstract{Survival models incorporating random effects to account for unmeasured
heterogeneity are being increasingly used in biostatistical and applied
research. Specifically, unmeasured covariates whose lack of inclusion in
the model would lead to biased, inefficient results are commonly
modelled by including a subject-specific (or cluster-specific) frailty
term that follows a given distribution (e.g.~Gamma or log-Normal).
Despite that, in the context of parametric frailty models little is
known about the impact of misspecifying the baseline hazard, the frailty
distribution, or both. Therefore, our aim is to quantify the impact of
such misspecification in a wide variety of clinically plausible
scenarios via Monte Carlo simulation, using open source software readily
available to applied researchers. We generate clustered survival data
assuming various baseline hazard functions, including mixture
distributions with turning points, and assess the impact of sample size,
variance of the frailty, baseline hazard function, and frailty
distribution. Models compared include standard parametric distributions
and more flexible spline-based approaches; we also included
semiparametric Cox models. The resulting bias can be clinically
relevant. In conclusion, we highlight the importance of fitting models
that are flexible enough and the importance of assessing model fit. We
illustrate our conclusions with two applications using data on diabetic
retinopathy and bladder cancer. Our results show the importance of
assessing model fit with respect to the baseline hazard function and the
distribution of the frailty: misspecifying the former leads to biased
relative and absolute risk estimates while misspecifying the latter
affects absolute risk estimates and measures of heterogeneity.}


\maketitle

\hypertarget{intro}{%
\section{Introduction}\label{intro}}

The standard, most common approach in medical research when dealing with
time to event data consists in fitting a Cox proportional hazards model,
where the baseline hazard is left unspecified and relative effect
estimates are frequently reported as the main quantities of interest.
However, it is often of interest to obtain absolute measures of risk: in
that context, modelling the baseline hazard has favourable properties,
and it can be achieved, for example, by using standard parametric
survival models with a simple parametric distribution (such as the
exponential, Weibull, or Gompertz distribution) or by using the flexible
parametric modelling approach of Royston and Parmar \citep{royston_2002}
to better capture the shape of complex hazard functions. The latter
approach requires choosing the number of degrees of freedom for the
spline term used to approximate the baseline hazard: in practice,
sensitivity analyses and information criteria (AIC, BIC) have been used
to select the best model. Recently, Rutherford et
al.~\citep{rutherford_2015} showed via simulation studies that, assuming
a sufficient number of degrees of freedom is used, the approximated
hazard function given by restricted cubic splines fits well for a number
of complex hazard shapes and the hazard ratios estimation is insensitive
to the correct specification of the baseline hazard.

Moreover, it is common to encounter clustered survival data where the
overall study population can be divided into heterogeneous clusters of
homogeneous observations. Examples are multi-centre clinical trial data,
individual-patient data meta-analysis, and observational data with
geographical clusters. With such data, the outcome variable is generally
recorded at the lowest hierarchical level while covariates can be
measured on units at any level of the hierarchy. As a consequence,
survival times of individuals within a cluster are likely to be
correlated and need to be analysed as such. Analogously, correlated data
may emerge as a consequence of recurrent events, i.e.~events that may
occur repeatedly within the same study subject. Unfortunately,
covariates that contribute to explaining the heterogeneity between
clusters are often not measured, e.g.~for economic reasons. Hence, the
frailty approach aims to account for the unobserved heterogeneity by
including a random effect that acts multiplicatively on the baseline
hazard and can be shared within a cluster.

Univariate frailty models have been first proposed by Vaupel \emph{et
al}. \citep{vaupel_1979} and Lancaster \citep{lancaster_1979}, and
further discussed by Hougaard \citep{hougaard_1984, hougaard_1986} with
specific focus on the frailty distribution. The Gamma distribution is
widely used, being mathematically very convenient; the inverse Gaussian
distribution is also common. A main difference between the two is that a
Gamma frailty yields a time-independent heterogeneity, while an inverse
Gaussian frailty yields heterogeneity that decays over time, making the
population more homogeneous as time goes by; in general, the relative
shapes of the individual and population hazard functions could differ
greatly because of the frailty effect. Additionally, Hougaard presents a
family of distributions with infinite mean, such as the reciprocal Gamma
distribution and the positive stable distribution. It is possible to use
log-Normal frailty as well; however, that leads to analytically
intractable formulae and additional computational complexity
(e.g.~requiring numerical quadrature or stochastic integration).
Hougaard also extended the univariate frailty approach to accommodate
frailty terms shared within cluster
\citep{hougaard_1986a, gutierrez_2002}, which results more attractive
when considering repeated event-times or clustered data that are
conditionally independent given the frailty \citep{hougaard_1995}.
Rondeau \emph{et al}. further extended the shared frailty model allowing
for hierarchical clustering of the data by including two nested frailty
terms \citep{rondeau_2006}, by allowing to study both heterogeneity
across trials and treatment-by-trial heterogeneity via additive frailty
models \citep{rondeau_2008}, and by jointly modelling recurrent events
and a dependent terminal event to jointly study the evolution of the two
processes or account for violations of the proportional hazard
assumption \citep{rondeau_2007, rondeau_2015, mazroui_2012}. Most of
these methods assume independence of the frailty terms; Ha \emph{et al}.
\citep{ha_2011} relaxed that assumption by developing frailty models
that can incorporate correlated frailty effects and/or
individual-specific frailty terms within the h-likelihood framework.
Crowther et al.~\citep{crowther_2014} generalised the frailty approach
by allowing for the inclusion of any number of random effects in a
parametric or flexible parametric survival model; under that framework,
a survival model with a shared log-Normal frailty term can be seen as a
survival model with a random intercept, shared between individuals that
belong to the same cluster. It follows that, under a more general
formulation, a mixed effects survival model can include not only a
random intercept (in which case it is equivalent to a model with a
log-Normal frailty) but multiple and potentially correlated random
effects as well. The semiparametric Cox model can be extended to
accommodate multilevel hierarchical structures as well; more information
and comparisons between different methods are presented in a recent
review by Austin \citep{austin_2017}.

As mentioned before, flexible parametric survival models are a robust
alternative to standard parametric survival models when the shape of the
hazard function is complex; using a sufficient number of degrees of
freedom, e.g.~2 or more, the spline-based approach is able to capture
the underlying shape of the hazard function with minimal bias. AIC and
BIC can guide the choice of the best fitting model, but they tend to
agree to within 1 or 2 degrees of freedom in practice
\citep{rutherford_2015}. Analogously, the impact of the choice of a
particular parametric frailty distribution on the estimation and testing
of regression coefficients is minimal. Pickles and Crouchley
\citep{pickles_1995} showed how the estimated values and the
distribution of the likelihood ratio test statistic do not differ much
comparing a variety of models such as the Weibull survival model with a
Gamma or log-Normal frailty. They conclude by arguing that convenience
and generality of the baseline hazard would seem more important that
generality of the frailty distribution when fitting a frailty model.
Glidden and Vittinghoff reached the same conclusions, highlighting how
different frailty distributions can lead to appreciably different
association structures despite not greatly affecting the estimation of
regression coefficients \citep{glidden_2004}. Lee and Thompson
\citep{lee_2008} showed how violations of the normality assumption for
random effects in hierarchical models do not affect fixed effects
substantially while having a substantial impact on inference regarding
the random effects. Thus, they advocate the use of more flexible
distributions such as the t or the skewed t for the random effects when
the distribution of the random effects is of interest - e.g.~in the
context of meta-analysis - despite the increased complexity. Liu
\emph{et al}. \citep{liu_2017} showed that flexible parametric survival
models perform well, both in terms of estimating the regression
coefficients and the variance of the frailty; in comparison,
semiparametric frailty models with a log-Normal frailty underestimated
the variance of the frailty. Liu \emph{et al}. \citep{liu_2017} also
showed that model misspecification could lead to an inflated estimated
variance of the frailty and a biased estimated survival function.
Finally, Ha \emph{et al}. showed (in the h-likelihood framework) that
misspecifying the baseline hazards results in larger bias than assuming
the wrong frailty distribution \citep{ha_2003}. In conclusion, the small
impact of misspecifying the frailty distribution on regression
coefficients seems to be well established in the literature, with some
evidence pointing towards biased absolute measures of risk.
Nevertheless, the structure of the frailty can be as important as the
baseline hazard choice as it gets easier to distinguish between
unobserved heterogeneity and non-proportional hazards as more
information on the correlation structure is available
\citep{elbers_1982, balan_2018}. However, little is known about the
impact of misspecifying the baseline hazard in survival models with
frailty terms. With this work, we aim therefore to assess the impact of
misspecifying the baseline hazard, the distribution of the frailty, or
both on measures of relative (regression coefficients) and absolute
(loss in life expectancy \citep{andersson_2013}) risk, and on
heterogeneity measures such as the estimated variance of the frailty
component. We compare a large set of models under different data
generating mechanisms: semiparametric and fully parametric survival
models with frailties, models with flexible baseline hazard, and models
with flexible baseline hazard and a penalty for the complexity of the
spline. Duchateau \emph{et al}. \citep{duchateau_2002} showed via
simulation studies that the number of centres and the number of patients
per centre influence the quality of the estimates, and they argue (in
the context of multi-centre clinical trials) the importance of making
sure that a trial is sufficiently large for the estimated heterogeneity
parameter to actually describe the true heterogeneity between centres
and not just random variability. Our data-generating mechanisms will
include this complexity and will cover combinations of number of
clusters and number of individuals per cluster. In addition to that, we
use readily available, open source software that practitioners and
applied biostatisticians can use in their analyses.

The models included in this comparison are fit to two applied settings.
First, we use a dataset consisting of a random sample of high-risk
patients from the Diabetic Retinopathy Study (DRS)
\citep{drs_1976, huster_1989} to illustrate our findings. The DRS was a
randomised, controlled clinical trial involving 15 clinical centres,
with a total of 1,758 patients enrolled between 1972 and 1975 and
follow-up until 1979. Each patient had one eye randomised to laser
treatment, while the other eye received no treatment. A failure occurred
when visual acuity dropped to below 5/200. The study outcome was time
between treatment and blindness, and censoring was caused by death,
dropout, or end of the study. Second, we use a subset of patients from a
full dataset of patients with bladder cancer constructed by Sylvester
\emph{et al}. \citep{eortc} by aggregating data from 7 trials organised
by the European Organization for Research and Treatment of Cancer
(EORTC). The trials compared several prophylactic treatments in bladder
cancer patients that we dichotomise for simplicity in
\emph{chemotherapy} vs \emph{no chemotherapy}; the outcome of interest,
in this case, is time to recurrence of bladder cancer after
randomisation to treatment. In our applications we focus on using the
AIC and BIC to select the functional form of the baseline hazard and the
frailty distribution that best fit the available data. Of course,
traditional tests based e.g.~on Martingale and Schoenfeld residuals are
also useful to fully test the correct specification of the model.
Further to that, novel tests are being developed e.g.~to test the
assumption of independence between censoring and the frailty
\citep{balan_2016}. These additional tests are outside the scope of our
manuscript, as we focus on specification of the baseline hazard and
frailty distribution - where selecting a specific functional form using
information criteria (such as AIC, BIC) is common practice.

The rest of this manuscript is laid out as follows. In Section
\ref{sim}, we introduce the simulation study and its aims,
data-generating mechanisms, estimands, methods, and performance measures
employed. In Section \ref{results}, we present the results of our
simulations. In Sections \ref{application-drs} and
\ref{application-bladder0} we compare the different models using real
data from a study on diabetic retinopathy and a study on bladder cancer,
respectively. Finally, we conclude the paper in Section \ref{discussion}
with a discussion.

\hypertarget{sim}{%
\section{A simulation study}\label{sim}}

\hypertarget{sim-aims}{%
\subsection{Aims}\label{sim-aims}}

The impact of misspecifying the baseline hazard or the frailty
distribution in survival models with shared frailty is not fully
understood. The primary aim of this simulation study consists in
assessing the consequences of such misspecification on estimates of
risk, both relative and absolute. This is particularly relevant as
parametric survival models are being increasingly used in applied
settings, with flexible frameworks and software readily available
\citep{crowther_2014a}. The advantages of parametric survival models are
well known: they allow easier absolute risk predictions and offer
advantages in terms of extrapolation and modelling of time-dependent
effects.

We simulate clustered survival data that we deemed clinically plausible,
as we aimed to mimic real data scenarios with each data-generating
mechanism: clustered studies such as multi-centre clinical trials,
individual patient data meta-analysis, paired organ studies, twin
studies, and so on.

\hypertarget{sim-dgms}{%
\subsection{Data-generating mechanisms}\label{sim-dgms}}

\begin{figure}

{\centering \includegraphics[width=0.4\textwidth]{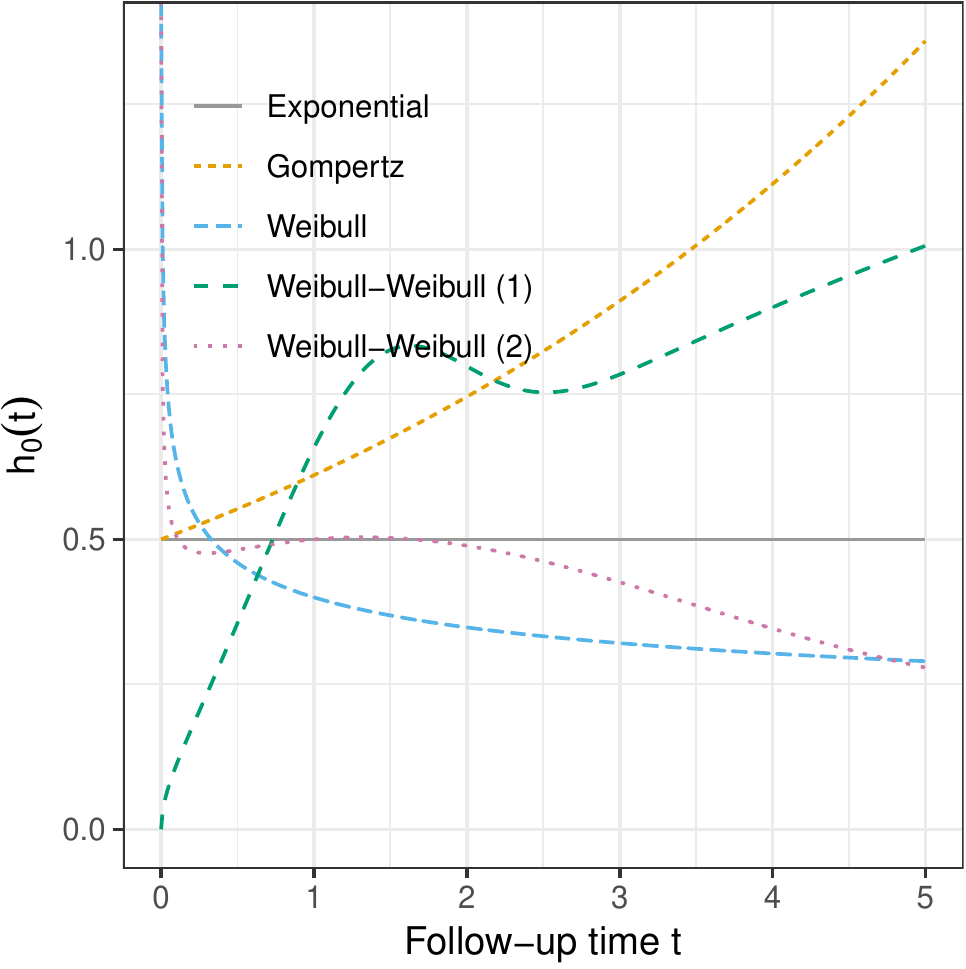}

}

\caption{Baseline hazard functions.}\label{fig:baseline-hazard-functions}
\end{figure}

We simulated data under five different baseline hazard functions using
the inversion method \citep{bender_2005} and its extension to
accommodate complex distributions with turning points
\citep{crowther_2013b}; specifically, we chose the exponential, Weibull,
Gompertz hazard functions, and two different two-components
Weibull-Weibull mixture distribution (Figure
\ref{fig:baseline-hazard-functions}, Table
\ref{tab:baseline-hazard-functions}). Then, for each baseline hazard
function, we simulated clustered data for 750 clusters of 2 individuals
each and for 20 clusters of 150 individuals each. We included a binary
treatment variable simulated from a Bernoulli random variable with
probability \(p = 0.50\) and an associated log-hazard ratio of \(-0.50\)
and cluster-specific frailty terms following either a Gamma or a
log-Normal distribution with variance \(\theta\),
\(\theta \in \{0.25, 0.75, 1.25\}\). Following a reviewer's suggestion,
we also investigated a mixture Normal frailty distribution. As a
motivation for this distribution, assume the presence of \(G = 2\)
hidden groups in each cluster (e.g.~an unmeasured binary covariate).
Formally, let \(g = \{1, 2\}\) be an index over the groups, with
\(\pi_g\) being the proportion in the groups and \(\sum_g \pi_g = 1\).
Let the hazard for the i\textsuperscript{th} individual in the
j\textsuperscript{th} cluster (and g\textsuperscript{th} hidden group)
be: \[
h(t | X_{ij}) = h_0(t) \exp(X_{gij} \beta + \sum_g \pi_g H_g),
\] where \(\sum_g \pi_g H_g\) follows a mixture Normal distribution with
mixing probabilities \(\pi_g\) and \(H_g \sim N(\mu_g, \sigma^2_g)\). We
assume \(\pi_1 = \pi_2 = 0.5\),
\(\mu = \{-3 \sqrt{\theta}, +3 \sqrt{\theta}\}\), and
\(\sigma^2_1 = \sigma^2_2 = \theta\) for the purposes of our
simulations.

Finally, we generated an event indicator variable by applying
administrative censoring at 5 years. In conclusion, we simulated
clustered survival data for 2 different sample sizes (number of
individuals and clusters), 3 possible distribution of the frailty
component, 3 frailty variances, and 5 baseline hazard functions: this
adds up to 90 different data-generating mechanisms.

\begin{table}
\caption{Parameters of data-generating baseline hazard functions}
\label{tab:baseline-hazard-functions}
\centering
\begin{tabular}{lc}
\toprule
Baseline hazard function & Parameters \\
\midrule
Exponential & $\lambda = 0.5$ \\
Weibull & $\lambda = 0.5, p = 0.8$ \\
Gompertz & $\lambda = 0.5, \gamma = 0.2$ \\
Weibull-Weibull (1) & $\lambda_1 = 0.3, \lambda_2 = 0.5, p_1 = 1.5, p_2 = 2.5, \pi = 0.7$ \\
Weibull-Weibull (2) & $\lambda_1 = 0.5, \lambda_2 = 0.5, p_1 = 1.3, p_2 = 0.7, \pi = 0.5$ \\
\bottomrule
\end{tabular}
\end{table}

\hypertarget{sim-estimands}{%
\subsection{Estimands}\label{sim-estimands}}

The estimands of interest are estimates of relative risk, absolute risk,
and heterogeneity. In addition, we monitor and report on convergence
rates of each model as well.

\paragraph{Relative risk}

The relative risk estimate of interest is the regression coefficient
\(\beta\) associated with the binary treatment; this coefficient can be
interpreted as the log-treatment effect, conditional on the unobserved
value of the frailty term. It is important to bear in mind therefore
that the hazard ratio in a frailty model carries the usual
interpretation only when comparing two hazards conditional on a given
frailty; unconditionally, at a population level, the proportionality of
hazards is not guaranteed to hold even under the proportional hazards
parametrisation. For most frailty distributions (including the Gamma and
log-Normal) the conditional hazard ratio is a true hazard ratio only at
time \(t = 0\), as the effect of the covariates on the hazard varies
over time depending on the actual distribution of the frailty
\citep{hougaard_1986, gutierrez_2002}.

\paragraph{Absolute risk}

The absolute risk estimate of interest is the 5-years loss in life
expectancy (LLE) (associated with the treatment of interest), defined as
the difference in life expectancy between exposed and non-exposed
individuals. The marginal 5-years life expectancy (LE) for exposed
individuals (\(X = 1\)) is defined as \[
\begin{aligned}
\text{LE}(X = 1) &= \int_0^5 S(u | X = 1) \ du \\
&= \int_0^5 \int_A S(u | X = 1, \alpha) \ p(\alpha) \ d \alpha \ du,
\end{aligned}
\] where \(A\) is the domain of the frailty \(\alpha\) and \(p(\alpha)\)
its density function. Consequently, the LLE associated with not being
exposed (\(X = 0\)) compared with being exposed is defined as \[
\text{LLE} = \int_0^5 S(u | X = 1) \ du - \int_0^5 S(u | X = 0) \ du.
\] The inner integral in LE (i.e.~\(S(t | X = 1)\)) has a closed form
when the frailty follows a Gamma distribution; with a log-Normal frailty
(and with a mixture Normal frailty), numerical integration is required.
We use the \texttt{quadinf} function from the \texttt{pracma} package in
R to perform numerical integration, which implements the double
exponential method for fast numerical integration of smooth real
functions on finite intervals \citep{takahasi_1974}. For infinite
intervals, the tanh-sinh quadrature scheme is applied
\citep{bailey_2006}. The outer integral in LE, however, is approximated
using spline-based integration as follows. We first estimate LE over
1,000 values of \(t\) between the minimum and the maximum observed
survival times; then, we fit an interpolating natural spline function
over the 1,000 LE estimates from step (1), which we finally integrate
between 0 and 5 (years) using the double exponential method of
\texttt{quadinf}. LLE follows by computing the difference between the
two integrals. Finally, we computed the standard error of the estimated
LLE using the numerical delta method (as implemented in the
\texttt{predictnl} function from the \texttt{rstpm2} R package).

\paragraph{Heterogeneity}

With this simulation study we mainly focus on estimates of risk.
Regardless, measures of heterogeneity are sometimes used to quantify
dependence between clustered observations. Therefore, we reported the
results of our simulations regarding the frailty variance in the
Supporting Web Material only. As the frailty variance estimated by
models assuming either a Gamma or a log-Normal distribution are not
directly comparable (being modelled on different scales, hazard versus
log-hazard), we do not include summary statistics for the frailty
variance where the frailty distribution is misspecified.

\hypertarget{sim-methods}{%
\subsection{Methods}\label{sim-methods}}

A general shared frailty model, assuming proportional hazards, has the
form \[
h(t_{ij} | X_{ij}) = \alpha_i h_0(t_{ij}) \exp(X_{ij} \beta)
\] when assuming a Gamma-distributed frailty with mean \(1\) and
variance \(\theta\). \(i\) indexes the cluster and \(j\) indexes the
individual. \(t_{ij}\) and \(X_{ij}\) are the survival time and
covariates of the \(j^\text{th}\) individual, \(i^\text{th}\) cluster,
respectively; \(h(\cdot)\) is the hazard function, \(h_0(\cdot)\) is the
baseline hazard function, \(\beta\) is a vector of regression
coefficients, and \(\alpha_i\) represent the frailty term. The frailty
term is assumed to be independent of covariates. Conversely, with a
log-Normal frailty, a general shared frailty model has the form \[
h(t_{ij} | X_{ij}) = h_0(t_{ij}) \exp(X_{ij} \beta + \eta_i),
\] with \(\exp(\eta_i) = \alpha_i\). \(\eta_i\) is assumed to have a
mean of \(0\) and a variance of \(\sigma ^ 2\). The latter model has a
convenient interpretation: \(\eta_i\) can be thought of as a random,
cluster-specific, intercept.

The conditional survival function from a frailty model is \[
S_{ij}(t | \alpha_i) = S_{ij}(t) ^ {\alpha_i}.
\] In this setting, the cluster-specific contribution to the likelihood
is obtained by calculating the cluster-specific likelihood conditional
on the frailty, consequently integrating out the frailty itself: \[
L_i = \int_A L_i(\alpha_i) f(\alpha_i) \, d\alpha,
\] with \(L_i(\alpha_i)\) the cluster-specific contribution to the
likelihood, conditional on the frailty. The cluster-specific
contribution to the likelihood is \[
L_i(\alpha_i) = \alpha_i ^ {D_i} \prod_{j = 1} ^ {n_i} S_{ij}(t_{ij}) ^ {\alpha_i} h_{ij}(t_{ij}) ^ {d_{ij}},
\] with \(D_i = \sum_{j = 1} ^ {n_i} d_{ij}\). A closed form formula for
\(L_i\) can be obtained when the frailty follows a Gamma distribution,
with further details provided elsewhere \citep{gutierrez_2002}.
Otherwise, assuming a log-Normal frailty, the likelihood is analytically
intractable and requires numerical integration to be performed (using
method such as adaptive Gauss-Hermite quadrature \citep{liu_1994}).
Consequently, it is important to bear in mind then that the performance
of the model fitting procedure will depend on the quality of the
approximation. For instance: methods that rely on Gaussian quadrature
will require an adequate number of integration points to provide an
accurate approximation, while Laplace approximation performs poorly with
a small number of clusters.

We compare a variety of different shared frailty models within this
simulation study. First, we fit semiparametric shared frailty models by
leaving the baseline hazard function \(h_0(t_{ij})\) unspecified and
assuming either a Gamma or a log-Normal frailty (Model 1-2, denoted by
\texttt{Cox} in plots). Second, we fit fully parametric survival models
by assuming that the baseline hazard function follows an exponential,
Weibull, or Gompertz distribution; we fit each of the three models
assuming both a Gamma and a log-Normal frailty distribution (Model 3-8,
denoted by \texttt{Exp}, \texttt{Wei}, and \texttt{Gom}). As a
comparison, we fit the Weibull model with both Gamma and log-Normal
frailties using the R package \texttt{frailtypack} as well (Model 17-18,
denoted by \texttt{FP(W)}). Third, we fit Royston-Parmar flexible
parametric survival models generalised by Liu \emph{et al}. to account
for clustered and correlated survival data
\citep{royston_2002, liu_2018, liu_2017}. Using the generalised survival
model formulation, the model is formulated as \[
S(t_{ij} | X_{ij}, \alpha_i) = \{G(\eta(t_{ij}, X_{ij}; \beta))\} ^ {\alpha_i},
\] with \(G(\cdot) = g^{-1}(\cdot)\) an inverse link function and
\(\eta(\cdot)\) a linear predictor function of time and covariates.
Choosing the log-log link function, the model is a proportional hazards
model; modelling the log of time with natural splines, we obtain a
Royston-Parmar model whose parameters can be fit using fully parametric
maximum likelihood. We assume 3, 5, or 9 degrees of freedom for the
natural spline of time and fit models using both Gamma and log-Normal
shared frailties (Model 9-14, denoted by \texttt{RP(df)} where
\texttt{df} is the number of degrees of freedom). In order to avoid
choosing the number of degrees of freedom for the spline term, we
estimate the same Royston-Parmar model with shared frailties using
penalised likelihood \citep{liu_2017} and either a Gamma or log-Normal
frailty (Model 15-16, denoted by \texttt{RP(P)}). The penalty term
accounts for the complexity of the smoother of time to avoid overfitting
the data; however, additional computational complexity is required to
select the smoothing parameter (or parameters). More details on the
penalised marginal likelihood estimation procedure are provided in Liu
\emph{et al}. \citep{liu_2017}. Finally, we fit shared frailty models
where the baseline hazard function is approximated by cubic M-splines on
the hazard scale \citep{rondeau_2012}; such models are fitted using
penalised likelihood, and we fix the smoothing parameter \(\kappa\) to
the value 10 and 10000 as in the simulations of Liu \emph{et al}.
\citep{liu_2017} (Model 19-22, denoted by \texttt{FP(k=kappa)}, with
\texttt{kappa} the smoothing parameter \(\kappa\)).

\hypertarget{sim-perfm}{%
\subsection{Performance measures}\label{sim-perfm}}

The first performance measure of interest is bias, quantifying whether
an estimator targets the true value on average. Formally, it is defined
as \(E(\hat{\beta}) - \beta\), with \(\hat{\beta}\) estimates of the
parameter \(\beta\). Second, we are interested in coverage, i.e.~the
proportion of times the \(100 \times (1 - \alpha)\%\) confidence
interval \(\hat{\beta} \pm Z_{1 - \alpha / 2} \times SE(\hat{\beta})\)
includes the true value \(\beta\). This allows assessing whether the
empirical coverage rate approaches the nominal coverage rate
(\(100 \times (1 - \alpha)\%\)). Finally, we are interested in mean
squared error {[}MSE{]}; MSE is the sum of the squared bias and variance
of \(\hat{\beta}\) and represents a natural way to integrate both
performance measures into one. However, the relative influence of bias
and variance of \(\hat{\beta}\) varies with the number of simulations
making generalising results difficult. Further details on the
performance measures of interest are given in Burton \emph{et al}.
\citep{burton_2006} and Morris \emph{et al}. \citep{morris_2019}. We
also report on convergence rates for each model, and we include Monte
Carlo standard errors for bias, coverage, and MSE to quantify the
uncertainty in estimating such performance measures
\citep{morris_2019, white_2010}.

In order to avoid the inflation of summary statistics caused by software
spuriously declaring convergence, we manually declared as non-converged
all the model fits that returned standardised point estimates or
standardised standard errors larger than 10 in absolute value. We
standardised values using median and inter-quartile range for
robustness.

\hypertarget{sim-numofsim}{%
\subsection{Number of simulations}\label{sim-numofsim}}

We generate \(1,000\) simulated data sets for each scenario; with
\(1,000\) replications, assuming a variance of \(0.1\) for the estimated
bias, we expect a Monte Carlo standard error of \(0.01\) for the
estimated bias. Being bias the key performance measure of interest, we
deemed this acceptable. Additionally, Monte Carlo standard error for
coverage is maximised when coverage is \(50\%\); with \(1,000\)
replications, the expected Monte Carlo standard error for coverage would
be \(1.58\%\). Should coverage be optimal at \(95\%\), the expected
Monte Carlo standard error would be \(0.68\%\). We deemed the expected
Monte Carlo standard error for coverage to be acceptable too.

\hypertarget{sim-software}{%
\subsection{Software}\label{sim-software}}

All models included and compared in this simulation study can be fitted
using R and standard, freely available user-written packages.
Furthermore, we did not tweak any of the convergence parameters utilised
for declaring convergence of the estimation algorithm. The
semiparametric shared frailty models can be fitted using the
\texttt{frailtyEM} and \texttt{coxme} packages for a Gamma and
log-Normal frailty, respectively. \texttt{frailtyEM} fits a Gamma
frailty model using the expectation-maximisation {[}EM{]} algorithm
\citep{dempster_1977}, while \texttt{coxme} relies on penalised
likelihood \citep{ripatti_2000}. A comparison of R packages for fitting
semiparametric frailty models is given in Hirsch and Wienke, 2012
\citep{hirsch_2012}; other possible packages that support semiparametric
shared frailty models (using different estimation algorithms) are
\texttt{frailtySurv} \citep{monaco_2018} and \texttt{frailtyHL}
\citep{frailtyHL}. The fully parametric shared frailty models can be
fitted using the \texttt{parfm} package for either a Gamma or a
log-Normal frailty; models are fit using full likelihood, and Laplace
integration is used when required \citep{munda_2012}. The Weibull shared
frailty model is also fit using the R package \texttt{frailtypack} for
comparison (denoted by \texttt{FP(W)}). The flexible parametric models
on the log-cumulative hazard scale with shared frailties can be fitted
using the \texttt{rstpm2} package. When specifying the number of degrees
of freedom for modelling the baseline hazard full likelihood is used,
otherwise \texttt{rstpm2} relies on penalised likelihood estimation. The
models with the baseline hazard function approximated by cubic M-splines
on the hazard scale are fitted using the R package \texttt{frailtypack}.

All packages but \texttt{coxme} returned a standard error for the
estimated frailty variance; therefore, when fitting a semiparametric
shared frailty model with a log-Normal frailty, we bootstrapped the
standard error of the frailty variance using 1,000 bootstrap
replications and resampling at the cluster level to preserve the
within-cluster correlation.

Finally, only \texttt{frailtyEM}, \texttt{rstpm2}, and
\texttt{frailtypack} provided a function to predict marginal survival;
for \texttt{coxme} and \texttt{parfm}, we manually wrote ad-hoc R
functions to estimate marginal survival, using numerical integration
when required.

All the R code utilised to simulate clustered survival data and fit each
model from this simulation study is freely and openly available online
on Github: \url{https://github.com/ellessenne/frailtymcsim}; we denote
the version of each package used for our simulations in Table
\ref{tab:software}.

\begin{table}
    \caption{Version of each R package used for our simulations}
    \label{tab:software}
    \centering
    \begin{tabular}{lc}
        \toprule
        R Package & Version \\
        \midrule
        \texttt{frailtyEM} & \texttt{0.8.8} \\
        \texttt{coxme} & \texttt{2.2-10} \\
        \texttt{parfm} & \texttt{2.7.6} \\
        \texttt{frailtypack} & \texttt{3.0.2.1} \\
        \texttt{rstpm2} & \texttt{1.4.5} \\
        \bottomrule
    \end{tabular}
\end{table}

\hypertarget{results}{%
\section{Results}\label{results}}

Among the 90 simulated scenarios, we select a subset of them to focus on
for conciseness; specifically, we select all scenarios with 20 clusters
of 150 individuals each. This adds up to 45 simulated scenarios, and
results for the remaining 45 scenarios (with 750 clusters of 2
individuals each) are included in the Supporting Web Material available
online from the aforementioned GitHub repository
(\url{https://github.com/ellessenne/frailtymcsim}).

The full results of the simulation study and the summary statistics can
be downloaded as well from the same GitHub repository, where we included
the full results tabulated by estimand and data-generating scenario and
additional plots. We recommend downloading the full dataset and
exploring results interactively using the web app INTEREST, available
online at \url{https://interest.shinyapps.io/interest/} and as a
stand-alone, offline package at
\url{https://github.com/ellessenne/interest}.

\hypertarget{convergence-rates}{%
\subsection{Convergence rates}\label{convergence-rates}}

\begin{figure}

{\centering \includegraphics[width=0.4\textwidth]{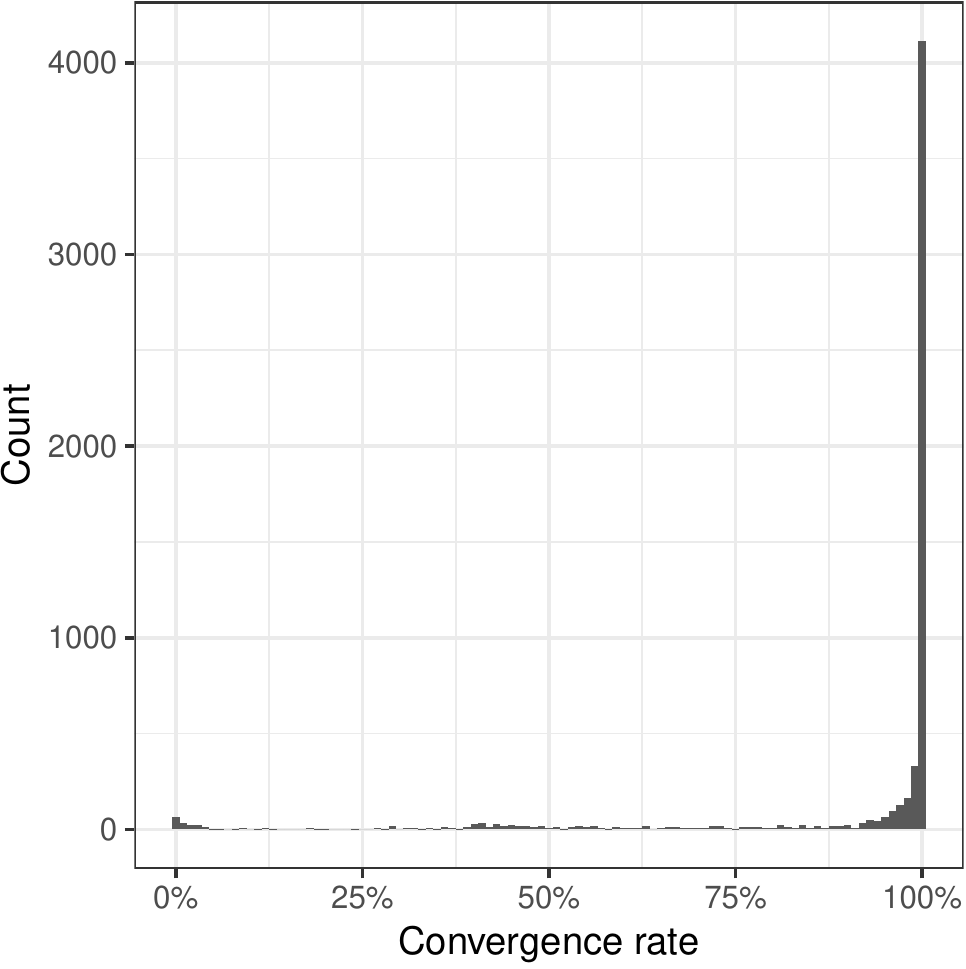}

}

\caption{Histogram of convergence rates per each model -- scenario combination.}\label{fig:convergence-histogram}
\end{figure}

Convergence rates were good for most models and most scenarios, with
75\% of model -- scenarios combinations showing a convergence rate of
98.5\% or above (Figure \ref{fig:convergence-histogram}). However, some
exceptions could be found:

\begin{enumerate}
\def\labelenumi{\arabic{enumi}.}
\tightlist
\item
  Parametric models with a Gompertz baseline hazard had the worst
  convergence rates, with a median convergence rate of 43.20\%
  (inter-quartile range: 29.25\% -- 55.85\%);
\item
  Parametric models with a Weibull baseline hazard and a log-Normal
  frailty fitted using the \texttt{frailtypack} package caused R to hang
  indefinitely in several scenarios, yielding a median convergence rate
  of 81.00\% (inter-quartile range: 65.30\% -- 97.30\%);
\item
  \texttt{frailtypack} models with a smooth baseline hazard modelled
  using M-splines showed low convergence rates for other scenarios as
  well, especially when simulating heterogeneity from a mixture Normal
  frailty distribution;
\item
  Some \texttt{frailtypack} models and some parametric models with a
  Gompertz baseline hazard did not converge at all in some scenarios
  with simulated data assuming a mixture Normal frailty;
\item
  All convergence rates are depicted in Online Web Figure 7, available
  in the Supporting Web Material.
\end{enumerate}

Furthermore, Online Web Figure 8 depicts predicted marginal
probabilities of non-convergence; the Gompertz models, the Weibull model
with a log-Normal frailty from \texttt{frailtypack}, and the model with
M-splines, a smoothing parameter \(k = 10000\), and a Gamma frailty
showed the highest predicted probabilities of non-convergence, all else
being equal. The probability of non-convergence also increased alongside
the variance of the frailty, and when simulating from a log-Normal or
mixture Normal (compared to a Gamma) distribution. Finally, Online Web
Figure 9 depicts the proportion of non-converged scenarios against the
average proportion of observed events per simulated scenario and model
included in our comparison. Models with worse convergence rates showed
an association between non-convergence and the average proportion of
events, with stronger right censoring being associated with worse
convergence rates. Ultimately, the factors that seemed to be associated
with convergence rates were the censoring proportion, the variance of
the frailty, and the distribution of the frailty. Nevertheless, the
software implementation and the algorithms used for fitting each model
seem to play an important role, with some software implementations being
more robust than others to variations in the factors outlined before.

\hypertarget{results-for-the-regression-coefficient}{%
\subsection{Results for the regression
coefficient}\label{results-for-the-regression-coefficient}}

Bias, coverage, and MSE of the regression coefficient for the scenarios
with 20 clusters of 150 individuals each are presented in Figures
\ref{fig:plot-scenario1-trt-bias},
\ref{fig:plot-scenario1-trt-coverage}, and
\ref{fig:plot-scenario1-trt-mse}. With a simple, exponential true
baseline hazard all models performed equally well in terms of bias and
coverage. Conversely, assuming a too simple parametric distribution with
a more complex true baseline hazard (or misspecifying the baseline
hazard) yielded a biased regression coefficient: significant positive
bias up to 0.175 and negative bias up to -0.105 for the models assuming
an exponential baseline hazard, and analogously for the models assuming
a Gompertz baseline hazard with positive bias up to 0.171 and negative
bias up to -0.112. A positive bias of 0.175 on the log-hazard ratio
scale corresponds to a 19\% relative risk overestimation; a negative
bias of -0.112 corresponds to a 11\% relative risk underestimation. The
semiparametric Cox models and all the flexible parametric models
(irrespectively of the number of degrees of freedom employed and of the
estimation procedure) yielded unbiased results, with the exception of
the model with 9 degrees of freedom when the true frailty followed a
mixture Normal distribution with component-specific variances of 1.25.
In that setting and assuming a Weibull and a mixture Weibull (2) true
baseline hazard the flexible parametric model yielded large biases -
although this seems to be a somewhat spurious result given the
performance of the same method in other scenarios. All models using
M-splines on the hazard scale performed similarly to the parametric
Weibull, with little to no bias; however, the performance of models
using M-splines worsened with a true mixture Normal frailty
distribution. Coverage was optimal for all models producing unbiased
estimates; however, coverage dropped considerably for models that
yielded biased estimates with coverage values as low as 5\% for models
showing the largest bias. Interestingly, misspecification of the frailty
distribution did not affect much the pattern of results; despite that,
bias seemed to worsen when we simulated the frailty from a log-Normal or
mixture Normal distribution compared to a Gamma distribution,
exacerbating the effect of misspecifying the baseline hazard. Finally,
the models that showed the lowest MSE were the semiparametric models,
the flexible parametric models, and the models using M-splines -
irrespectively of the true baseline hazard and distribution of the
frailty. The exponential and Gompertz parametric models showed a larger
MSE - up to 10-fold larger - when the baseline hazard was misspecified,
with the Gompertz model showing an MSE larger than semiparametric and
flexible parametric models even when well specified. The Weibull model,
as observed before, performed similarly to semiparametric and flexible
parametric models. Finally, this pattern of results was largely
maintained with the remaining scenarios (with 750 clusters of 2
individuals each); the corresponding plots are included in the
Supporting Web Material.

\hypertarget{results-for-the-5-years-lle}{%
\subsection{Results for the 5-years
LLE}\label{results-for-the-5-years-lle}}

Bias, coverage, and MSE for the 5-years LLE are presented in Figures
\ref{fig:plot-scenario1-lle-bias},
\ref{fig:plot-scenario1-lle-coverage}, and
\ref{fig:plot-scenario1-lle-mse}, respectively. The pattern we observed
for LLE mirrors the pattern observed for the regression coefficient:
models with a misspecified baseline hazard (or a baseline hazard not
flexible enough to capture the underlying shape) yielded biased results,
both positive and negative. Negative bias was up to -0.057 and positive
bias was up to 0.031: this corresponds, respectively, to a difference of
approximately (minus) 1 month and half a month in the estimated LLE. The
Weibull model fit with \texttt{frailtypack} largely underestimated the
5-years LLE in all scenarios (negative bias between -0.184 and -0.100),
and the M-splines model with smoothing parameter \(\kappa = 10000\) and
a Gamma distribution performed even worse when the true baseline hazard
followed a Weibull-Weibull (1) distribution (negative bias between
-0.461 and -0.169). The large bias observed for this M-splines model in
these scenarios seems to be spurious - analogously as before with the
flexible parametric model. Interestingly, models with a well-specified
frailty performed better than models with a misspecified frailty, both
for a true Gamma and log-Normal frailty. Conversely, when simulating
from a mixture Normal distribution all models yielded positively biased
results (up to 0.257, e.g.~3 months) with exceptions being the
\texttt{frailtypack} models described before, where underestimation of
the results still applied. Coverage followed a similar pattern, with
optimal coverage for models with small bias and reduced coverage for
models that yielded biased results; overall, coverage was better when
the frailty distribution was well specified. As a consequence of the
large positive bias, coverage in scenarios simulated from a mixture
Normal distribution was poor. Mean squared error was similar across all
scenarios with minimal variability and a notable exception: models
fitted using M-splines for the baseline hazard and a log-Normal frailty
had a much higher MSE, approximately 10 times larger. Another
interesting observation is that the empirical standard error (the
standard deviation of the estimated LLE) was systematically larger than
the mean of the standard errors for LLE obtained using the numerical
delta method (results included in the Supporting Web Material; we can
conclude that the numerical delta method used to obtain a standard error
for LLE underestimated the standard error. Finally, once again results
for the remaining scenarios (included in the Supporting Web Material)
are similar to the results presented in the main body of the manuscript.

\begin{landscape}

\begin{figure}

{\centering \includegraphics[width=1.33\textwidth]{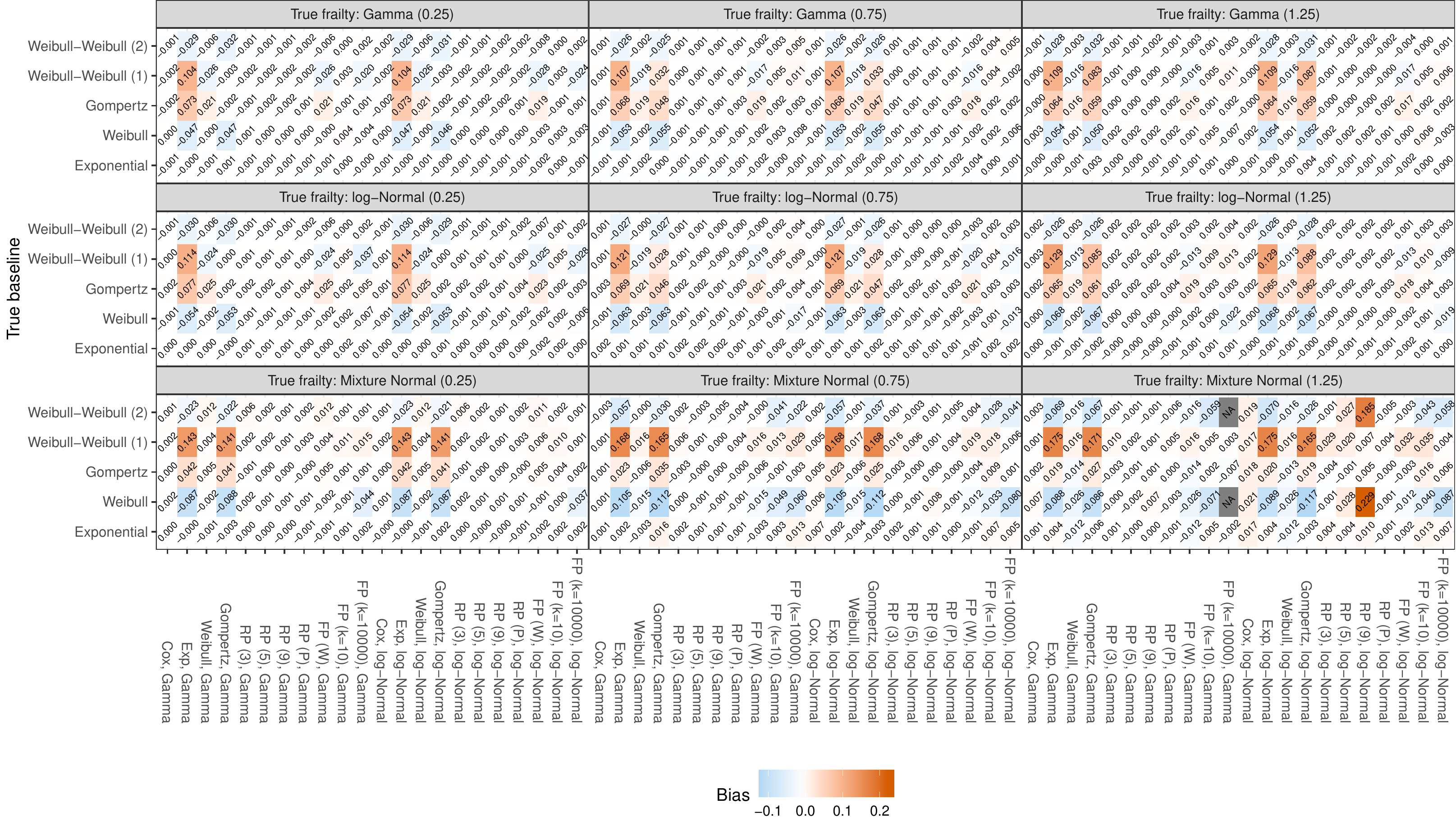}

}

\caption{Bias of regression coefficient, scenario: 20 clusters of 150 individuals each. Colours represent positive and negative bias, and solid grey represents scenarios where no model converged.}\label{fig:plot-scenario1-trt-bias}
\end{figure}

\begin{figure}

{\centering \includegraphics[width=1.33\textwidth]{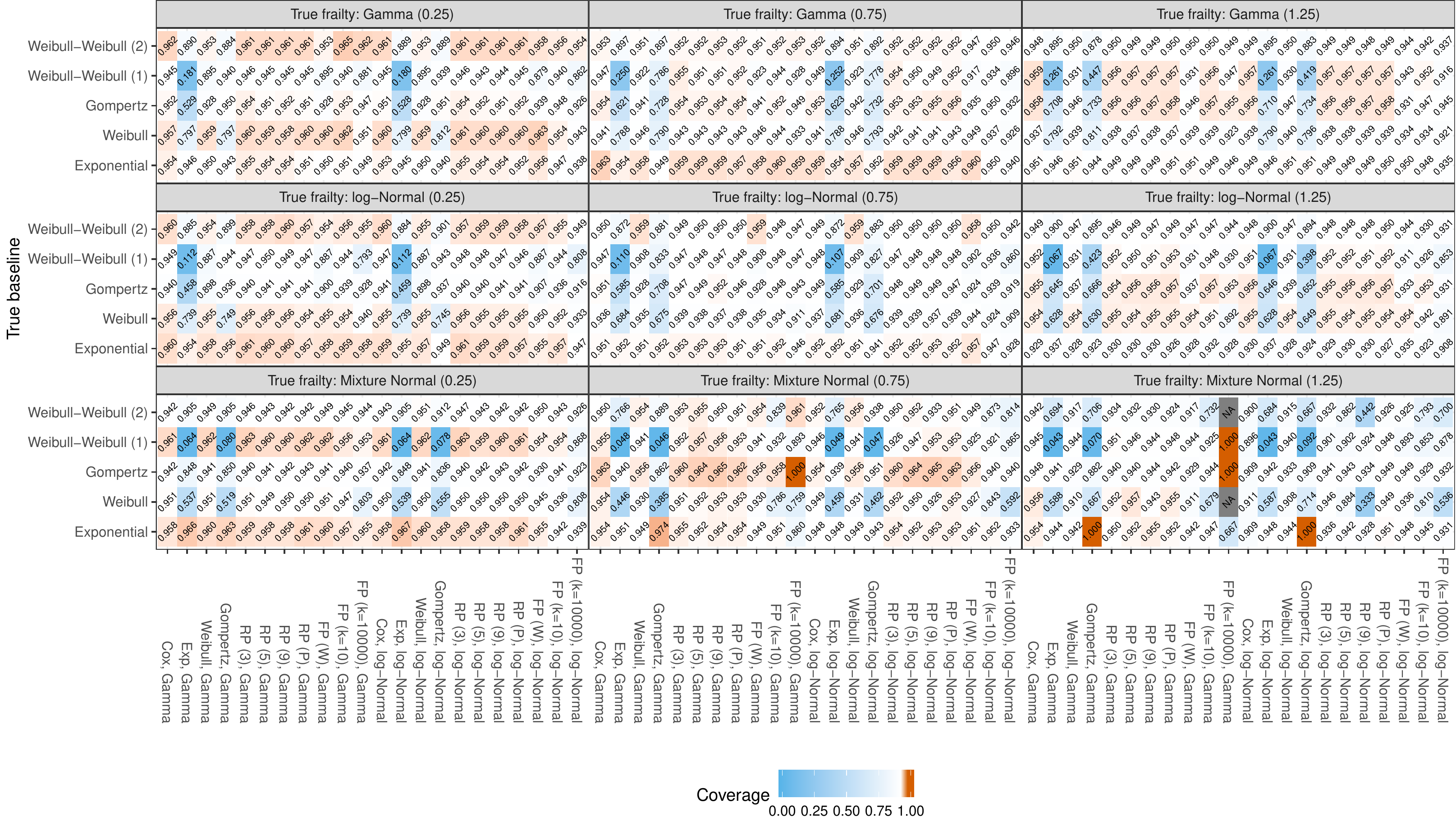}

}

\caption{Coverage of regression coefficient, scenario: 20 clusters of 150 individuals each. Colours represent positive and negative bias, and solid grey represents scenarios where no model converged.}\label{fig:plot-scenario1-trt-coverage}
\end{figure}

\begin{figure}

{\centering \includegraphics[width=1.33\textwidth]{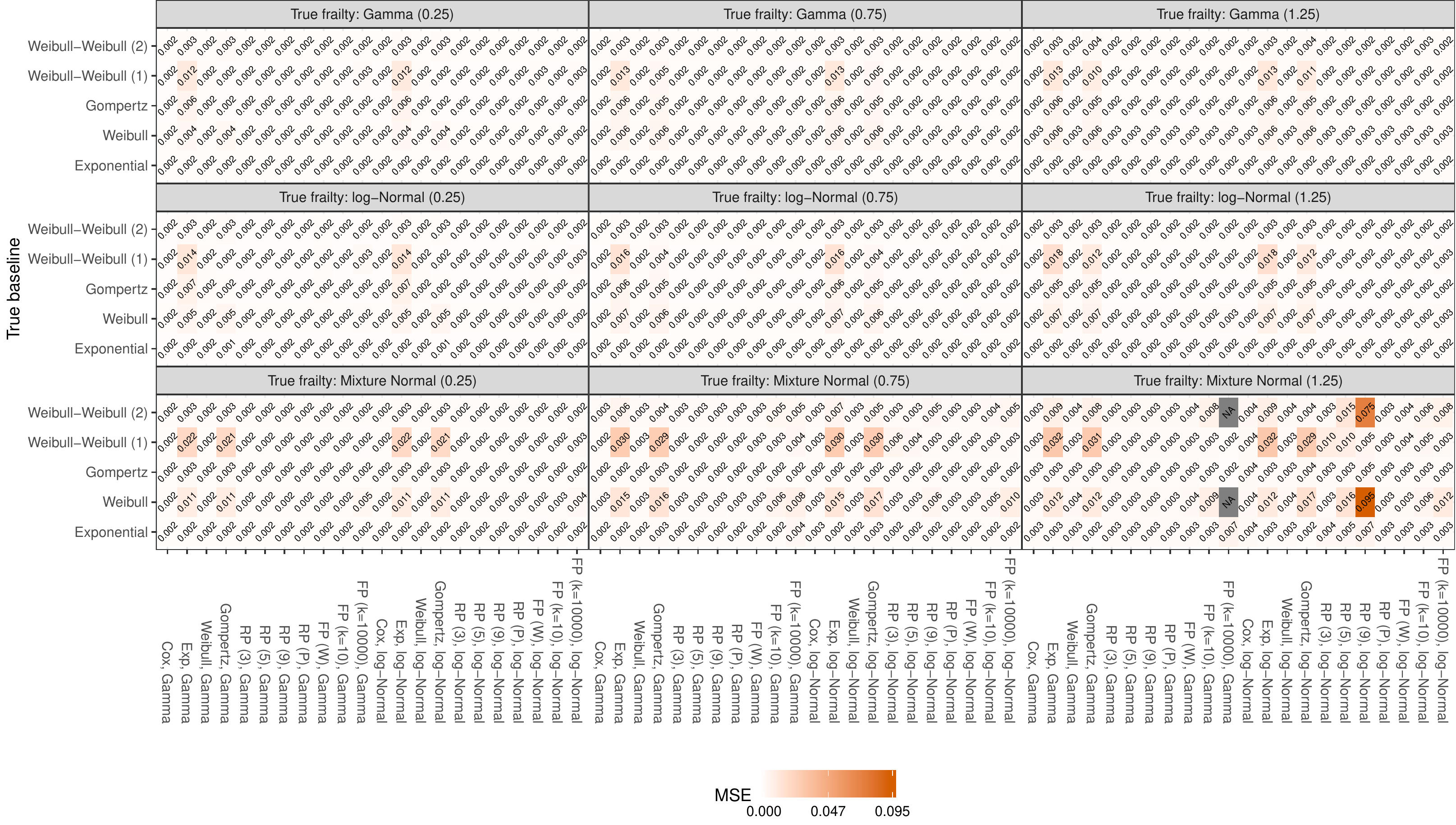}

}

\caption{Mean squared error of regression coefficient, scenario: 20 clusters of 150 individuals each. Colours represent positive and negative bias, and solid grey represents scenarios where no model converged.}\label{fig:plot-scenario1-trt-mse}
\end{figure}

\begin{figure}

{\centering \includegraphics[width=1.33\textwidth]{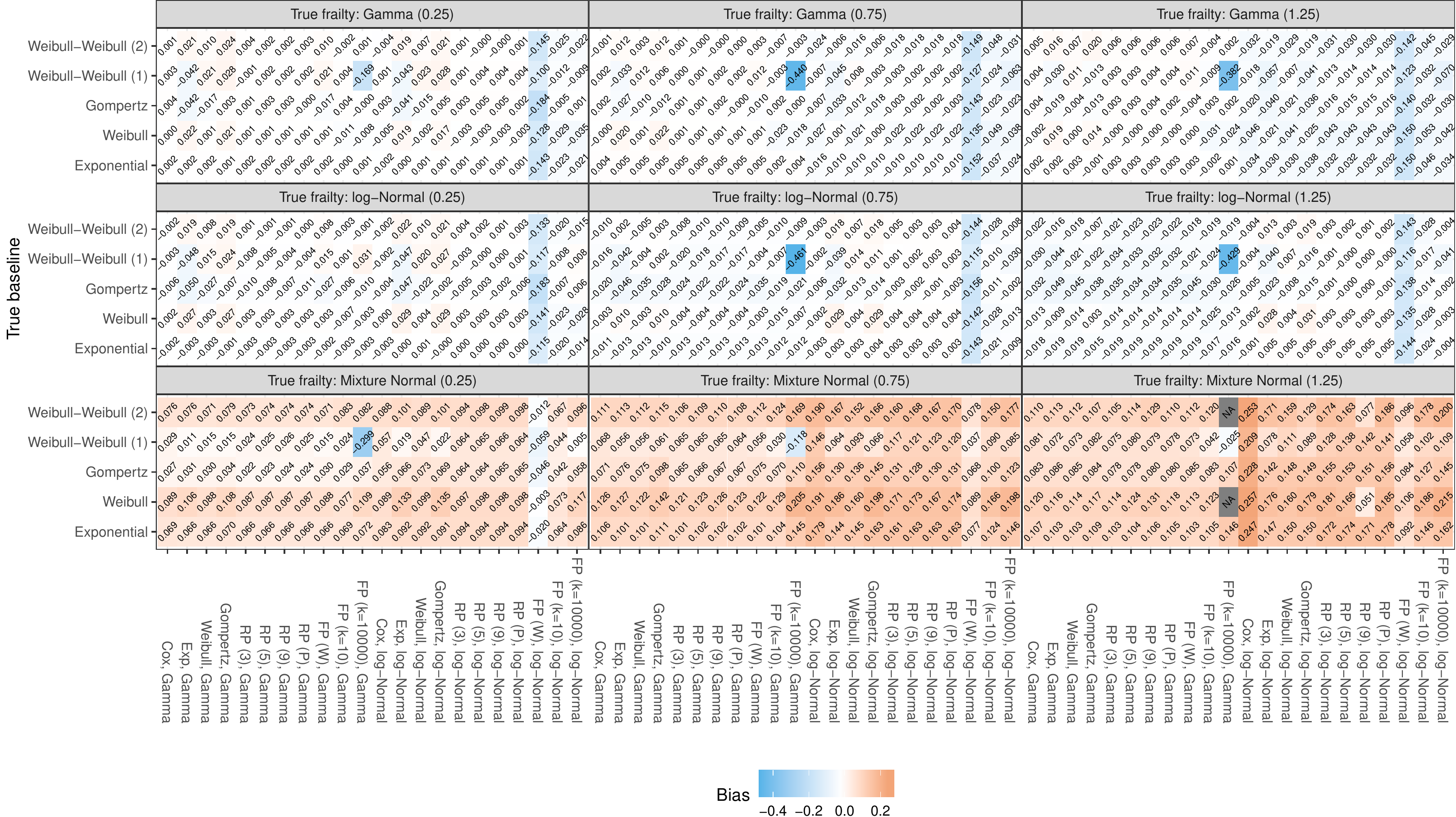}

}

\caption{Bias of LLE, scenario: 20 clusters of 150 individuals each. Colours represent positive and negative bias, and solid grey represents scenarios where no model converged.}\label{fig:plot-scenario1-lle-bias}
\end{figure}

\begin{figure}

{\centering \includegraphics[width=1.33\textwidth]{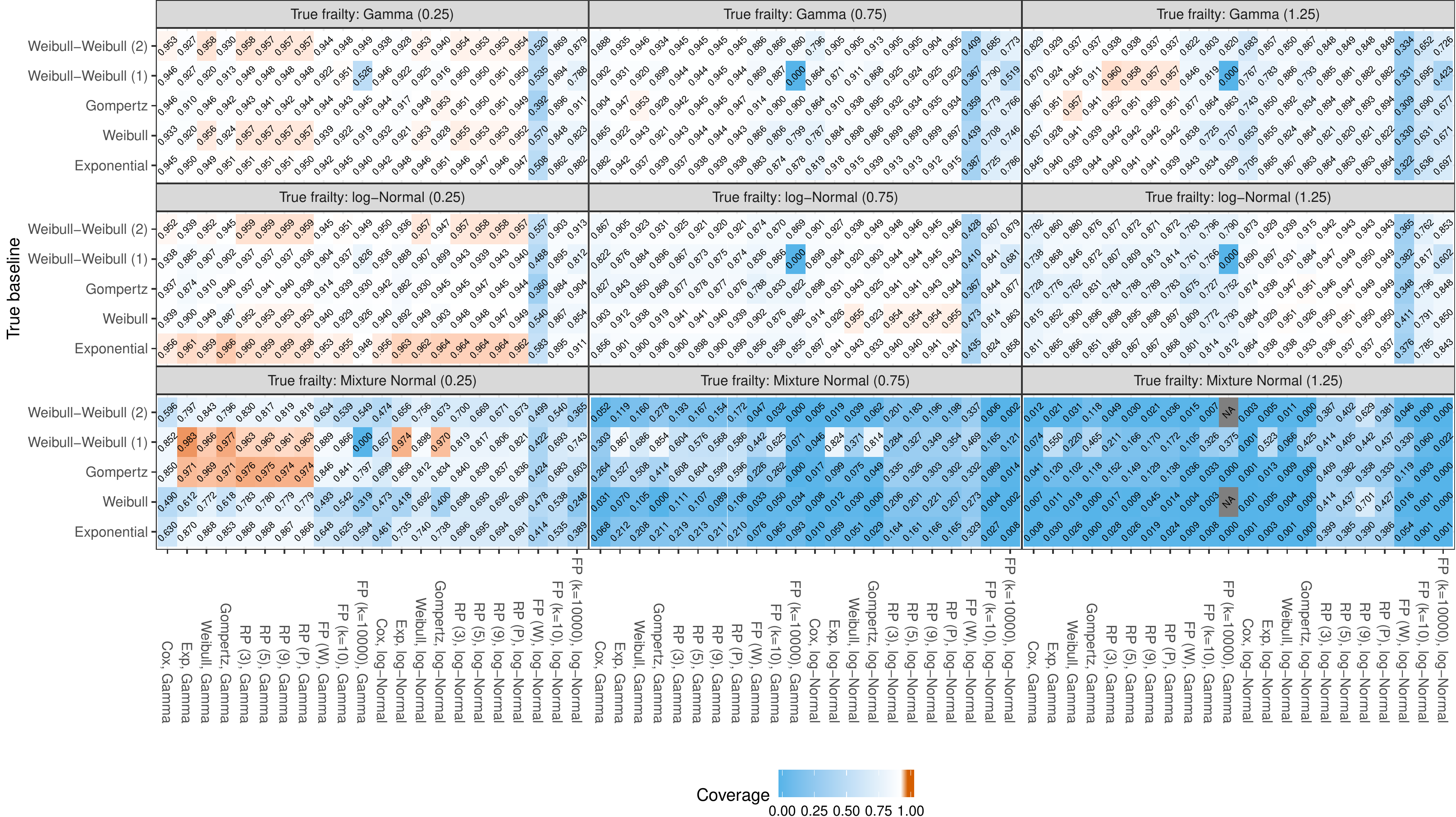}

}

\caption{Coverage of LLE, scenario: 20 clusters of 150 individuals each. Colours represent positive and negative bias, and solid grey represents scenarios where no model converged.}\label{fig:plot-scenario1-lle-coverage}
\end{figure}

\begin{figure}

{\centering \includegraphics[width=1.33\textwidth]{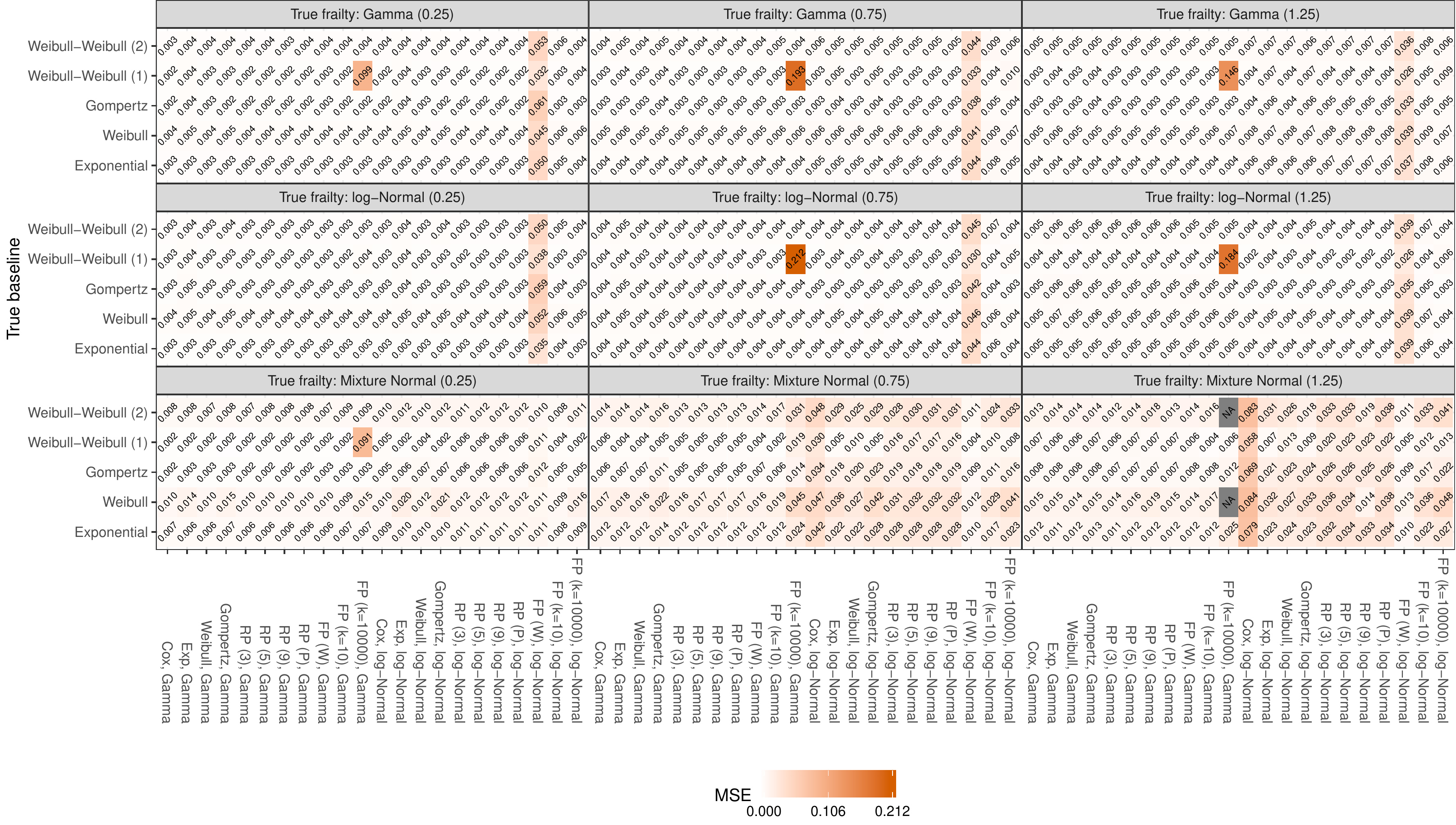}

}

\caption{Mean squared error of LLE, scenario: 20 clusters of 150 individuals each. Colours represent positive and negative bias, and solid grey represents scenarios where no model converged.}\label{fig:plot-scenario1-lle-mse}
\end{figure}

\end{landscape}

\hypertarget{application-drs}{%
\section{Application to diabetic retinopathy
data}\label{application-drs}}

In this section, we apply the models included in our simulation study to
a dataset from the Diabetic Retinopathy Study. The outcome of interest
here is time to blindness, and we include laser treatment as a binary
covariate. Laser treatment is the main exposure of interest, and we do
not include other covariates for simplicity. We define the subject ID as
the cluster indicator variable to account for correlation between eyes
of a given individual, and we aim to present estimates of treatment
effect and five-years LLE. The model we fit is a model of the kind: \[
h(t_{ij}) = \alpha_i h_0(t_{ij}) \exp(X_{ij} \beta),
\] with \(i\) and \(j\) being the individual-level and eye-level
indicator variables, \(t_{ij}\) the time to event for eye \(j\) of
individual \(i\), \(\beta\) being the treatment effect, \(X_{ij}\) the
treatment modality for eye \(j\) of individual \(i\), and \(\alpha_i\)
the frailty for individual \(i\). We model the baseline hazard
\(h_0(\cdot)\) via fully parametric and flexible parametric
distributions or by leaving it unspecified. The flexible parametric
models are modelled on the log-cumulative hazard scale: \[
\log H(t_{ij}) = s(\log(t_{ij}) | \gamma, k_0) + X_{ij} \beta + \eta_i,
\] with \(s(\cdot)\) a spline function of log-time with parameter vector
\(\gamma\) and knot vector \(k_0\). Despite being on the log-cumulative
hazard scale, the aforementioned model is still a proportional hazards
model. Finally, we model the frailty distribution assuming either a
Gamma or log-Normal distribution.

The dataset consisted of 394 observation clustered in 197 patients; the
median follow-up, estimated using the inverse Kaplan-Meier method
\citep{schemper_1996} and using robust standard errors to account for
clustering, was 51.10 months, with a confidence interval (based on the
estimated survival curve) of 47.60 to 54.37 months.

\begin{figure}

{\centering \includegraphics[width=0.8\textwidth]{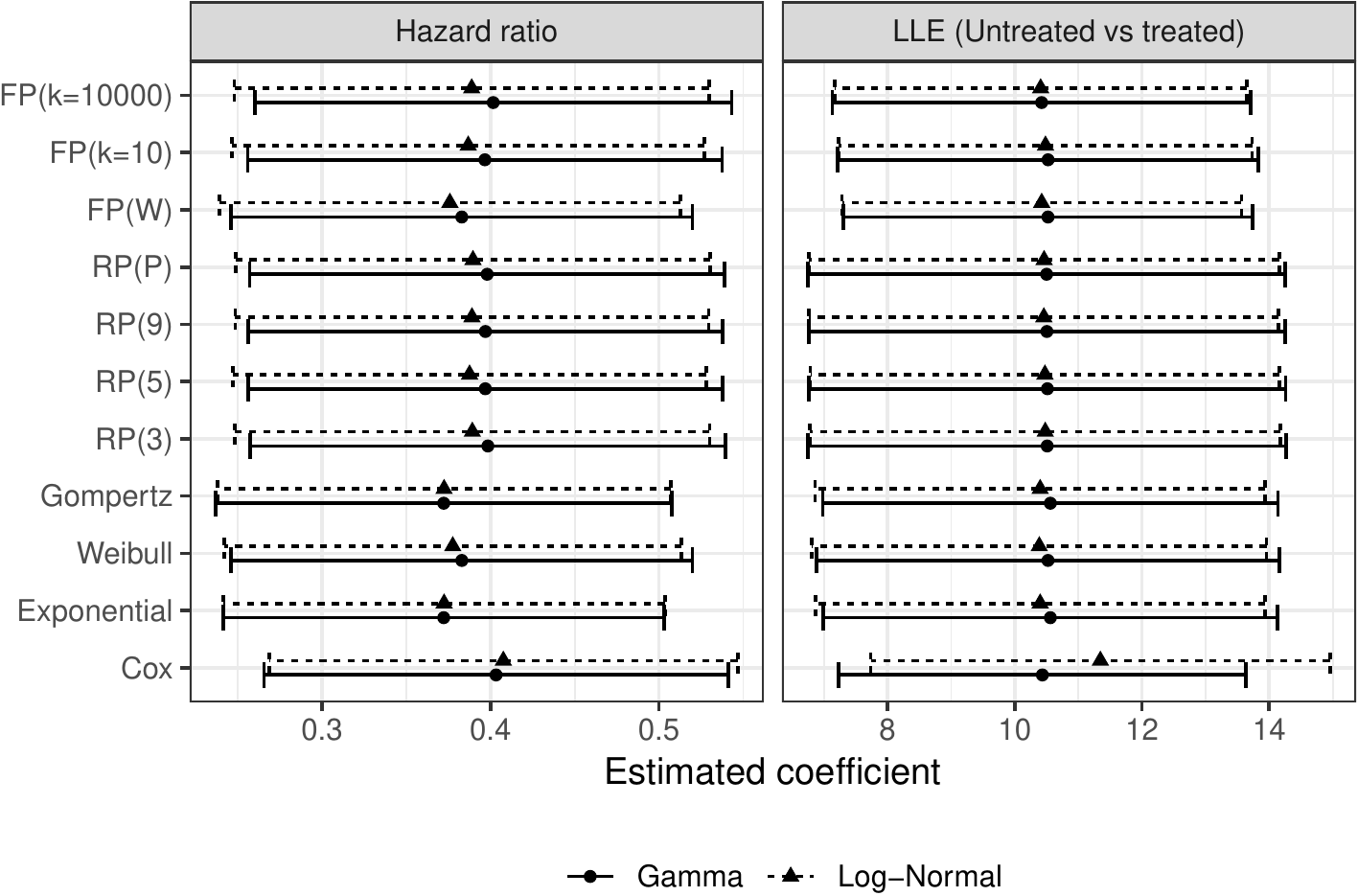}

}

\caption{Comparison of relative and absolute estimates of risk with confidence intervals, application to diabetic retinopathy data \label{fig:drs-forest-plot}}\label{fig:drs-forest-plots}
\end{figure}

\begin{table}[t]

\caption{\label{tab:drs-tab}Results: application of shared frailty models to a subset of data from the DRS. Values in rounded brackets are standard errors for each estimate.}
\centering
\begin{tabular}{rccc}
\toprule
Baseline & Hazard ratio & Frailty variance & LLE (Untreated vs treated)\\
\midrule
\addlinespace[0.3em]
\multicolumn{4}{l}{\textbf{Gamma frailty:}}\\
\hspace{1em}Cox & 0.4033 (0.0703) & 0.8477 (0.3140) & 10.4346 (1.6335)\\
\hspace{1em}Exponential & 0.3723 (0.0667) & 1.1490 (0.3028) & 10.5606 (1.8244)\\
\hspace{1em}Weibull & 0.3830 (0.0698) & 1.0402 (0.3293) & 10.5231 (1.8585)\\
\hspace{1em}Gompertz & 0.3723 (0.0691) & 1.1490 (0.3516) & 10.5606 (1.8253)\\
\hspace{1em}RP(3) & 0.3985 (0.0720) & 0.8875 (0.3185) & 10.5104 (1.9171)\\
\hspace{1em}RP(5) & 0.3969 (0.0718) & 0.9067 (0.3215) & 10.5140 (1.9117)\\
\hspace{1em}RP(9) & 0.3970 (0.0718) & 0.9046 (0.3208) & 10.5057 (1.9104)\\
\hspace{1em}RP(P) & 0.3981 (0.0720) & 0.8943 (0.3195) & 10.5016 (1.9138)\\
\hspace{1em}FP(W) & 0.3830 (0.0698) & 1.0402 (0.3294) & 10.5232 (1.6393)\\
\hspace{1em}FP(k=10) & 0.3967 (0.0719) & 0.9036 (0.3217) & 10.5241 (1.6857)\\
\hspace{1em}FP(k=10000) & 0.4016 (0.0722) & 0.8751 (0.3146) & 10.4233 (1.6790)\\
\addlinespace[0.3em]
\multicolumn{4}{l}{\textbf{Log-normal frailty:}}\\
\hspace{1em}Cox & 0.4076 (0.0709) & 0.7771 (0.2990) & 11.3468 (1.8417)\\
\hspace{1em}Exponential & 0.3725 (0.0669) & 1.2183 (0.3054) & 10.4002 (1.8043)\\
\hspace{1em}Weibull & 0.3776 (0.0693) & 1.1594 (0.3379) & 10.3862 (1.8232)\\
\hspace{1em}Gompertz & 0.3725 (0.0687) & 1.2183 (0.3487) & 10.4001 (1.8046)\\
\hspace{1em}RP(3) & 0.3892 (0.0719) & 1.0287 (0.3861) & 10.4800 (1.8889)\\
\hspace{1em}RP(5) & 0.3876 (0.0717) & 1.0560 (0.3916) & 10.4747 (1.8818)\\
\hspace{1em}RP(9) & 0.3890 (0.0717) & 1.0362 (0.3849) & 10.4579 (1.8833)\\
\hspace{1em}RP(P) & 0.3896 (0.0718) & 1.0270 (0.3839) & 10.4617 (1.8863)\\
\hspace{1em}FP(W) & 0.3759 (0.0697) & 1.1967 (0.4023) & 10.4241 (1.6023)\\
\hspace{1em}FP(k=10) & 0.3868 (0.0716) & 1.0621 (0.3919) & 10.4838 (1.6575)\\
\hspace{1em}FP(k=10000) & 0.3889 (0.0719) & 1.0623 (0.3934) & 10.4080 (1.6536)\\
\bottomrule
\end{tabular}
\end{table}

We first fit all the models without including any other covariate than
the exposure to treatment; results are presented in Figure
\ref{fig:drs-forest-plot} and Table \ref{tab:drs-tab}. The log-treatment
effect varied between -0.9882 and -0.8974, corresponding to hazard
ratios of 0.3723 to 0.4076, respectively. Analogously, the estimates of
5-years LLE ranged between 10.3862 and 11.3468 months for non-exposed
compared to exposed. LLE can be interpreted as follows: on average,
individuals not treated with laser experienced blindness approximately
11 months before individuals that received the treatment. The difference
in estimated risk (relative and absolute) appears to be relevant,
highlighting once again the importance of choosing an appropriate
functional form of the baseline hazard.

\begin{table}[t]

\caption{\label{tab:drs-tab-AIC-BIC}Results: comparison of AIC/BIC from shared frailty models, application to a subset of data from the DRS. Best AIC/BIC values are in bold.}
\centering
\begin{tabular}{rcc}
\toprule
Baseline & AIC & BIC\\
\midrule
\addlinespace[0.3em]
\multicolumn{3}{l}{\textbf{Gamma frailty:}}\\
\hspace{1em}Cox & --- & \vphantom{1} ---\\
\hspace{1em}Exponential & 1,661.96 & 1,673.89\\
\hspace{1em}Weibull & 1,663.49 & 1,679.39\\
\hspace{1em}Gompertz & 1,663.96 & 1,679.87\\
\hspace{1em}RP(3) & 1,663.11 & 1,686.97\\
\hspace{1em}RP(5) & 1,665.08 & 1,696.90\\
\hspace{1em}RP(9) & 1,659.40 & 1,707.12\\
\hspace{1em}RP(P) & --- & \vphantom{1} ---\\
\hspace{1em}FP(W) & 1,663.49 & 1,679.39\\
\hspace{1em}FP(k=10) & --- & \vphantom{1} ---\\
\hspace{1em}FP(k=10000) & --- & \vphantom{1} ---\\
\addlinespace[0.3em]
\multicolumn{3}{l}{\textbf{Log-normal frailty:}}\\
\hspace{1em}Cox & --- & ---\\
\hspace{1em}Exponential & \textbf{1,656.99} & \textbf{1,668.92}\\
\hspace{1em}Weibull & 1,658.87 & 1,674.78\\
\hspace{1em}Gompertz & 1,658.99 & 1,674.89\\
\hspace{1em}RP(3) & 1,662.01 & 1,685.87\\
\hspace{1em}RP(5) & 1,663.90 & 1,695.71\\
\hspace{1em}RP(9) & 1,658.37 & 1,706.09\\
\hspace{1em}RP(P) & --- & ---\\
\hspace{1em}FP(W) & 1,662.27 & 1,678.17\\
\hspace{1em}FP(k=10) & --- & ---\\
\hspace{1em}FP(k=10000) & --- & ---\\
\bottomrule
\end{tabular}
\end{table}

\begin{figure}

{\centering \includegraphics[width=0.4\textwidth]{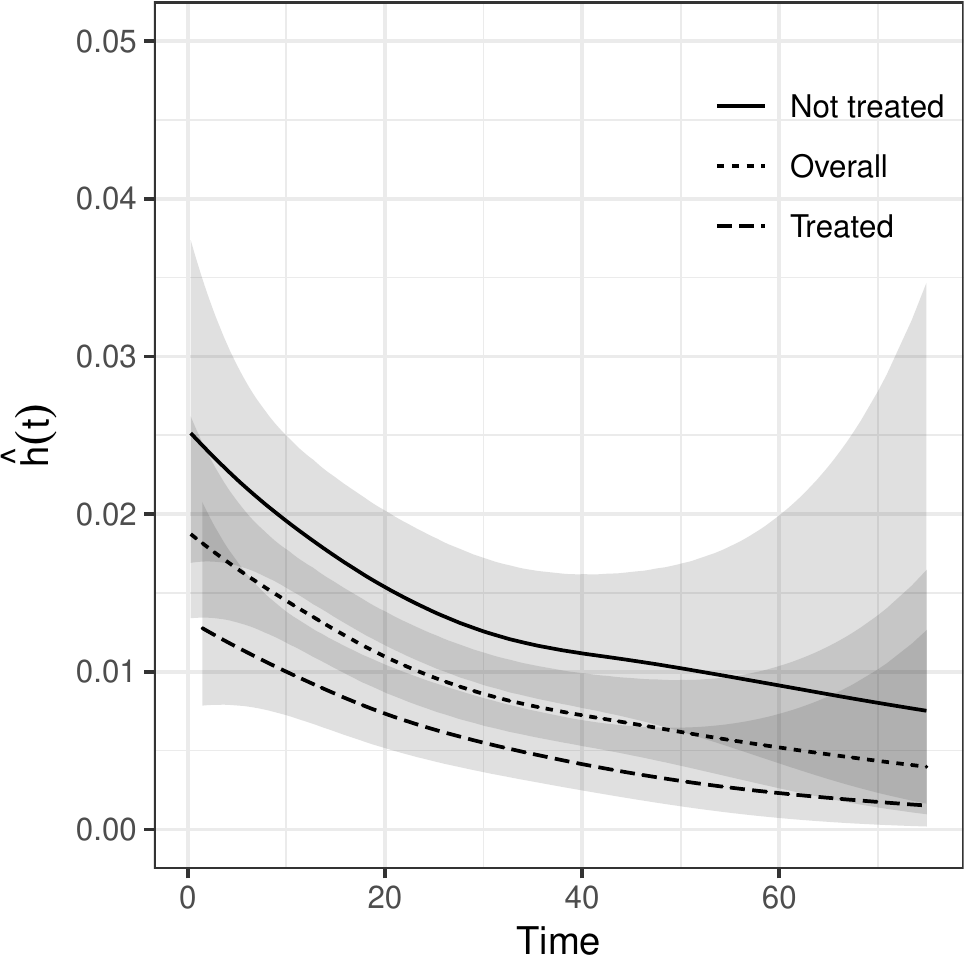}

}

\caption{Smooth non-parametric hazard estimate, overall and by treatment modality.}\label{fig:drs-est-hazard}
\end{figure}

In Table \ref{tab:drs-tab-AIC-BIC} we present AIC and BIC for the models
fitted using full likelihood on DRS data; the best model according to
each criterion is highlighted in bold. We do not include models fitted
using either partial or penalised likelihood in this comparison. The
best model according to both the AIC and the BIC is the model with a
log-Normal frailty and an exponential baseline hazard; however, a
non-parametric\sout{al} estimate of the hazard function using a
kernel-based method (Figure \ref{fig:drs-est-hazard}, R package
\texttt{bshazard}) seems to suggest against assuming an exponential
baseline hazard function. Therefore, we select the second-best model in
terms of AIC as the model we focus on, that is, the flexible parametric
model with the baseline hazard modelled via a restricted cubic spline
with 9 degrees of freedom and a log-Normal frailty. The penalised model
with a flexible parametric baseline hazard and a log-Normal frailty
selected an effective number of degrees of freedom of 9.6261, very
similar to the final model we selected. However, it is important to
remember that other information criteria are available, especially for
models with random effects. Vaida and Blanchard \citep{vaida_2005}
suggested the use of conditional AIC (cAIC) for model selection in
linear mixed models. They demonstrated that a classical AIC
(i.e.~marginal AIC) and its small sample correction are inappropriate
when the interest is on cluster: see also Liang, Wu and Zou
\citep{liang_2008}. Unfortunately, the cAIC is not routinely reported by
software fitting shared frailty models (with the exception of
\texttt{frailtyHL} \citep{ha_2012, frailtyHL}). We encourage the
maintainers of such R packages to add it to their packages, enabling
applied researchers to use information criteria for model selection that
are more appropriate in the settings of random effects models. Using the
flexible parametric approach, it is straightforward to model the
baseline hazard function in continuous time and it is possible to obtain
smooth predictions. It is also straightforward to accommodate
time-varying covariates effects (and therefore assess the proportional
hazards assumption), as we demonstrate next.

\begin{figure}

{\centering \includegraphics[width=0.8\textwidth]{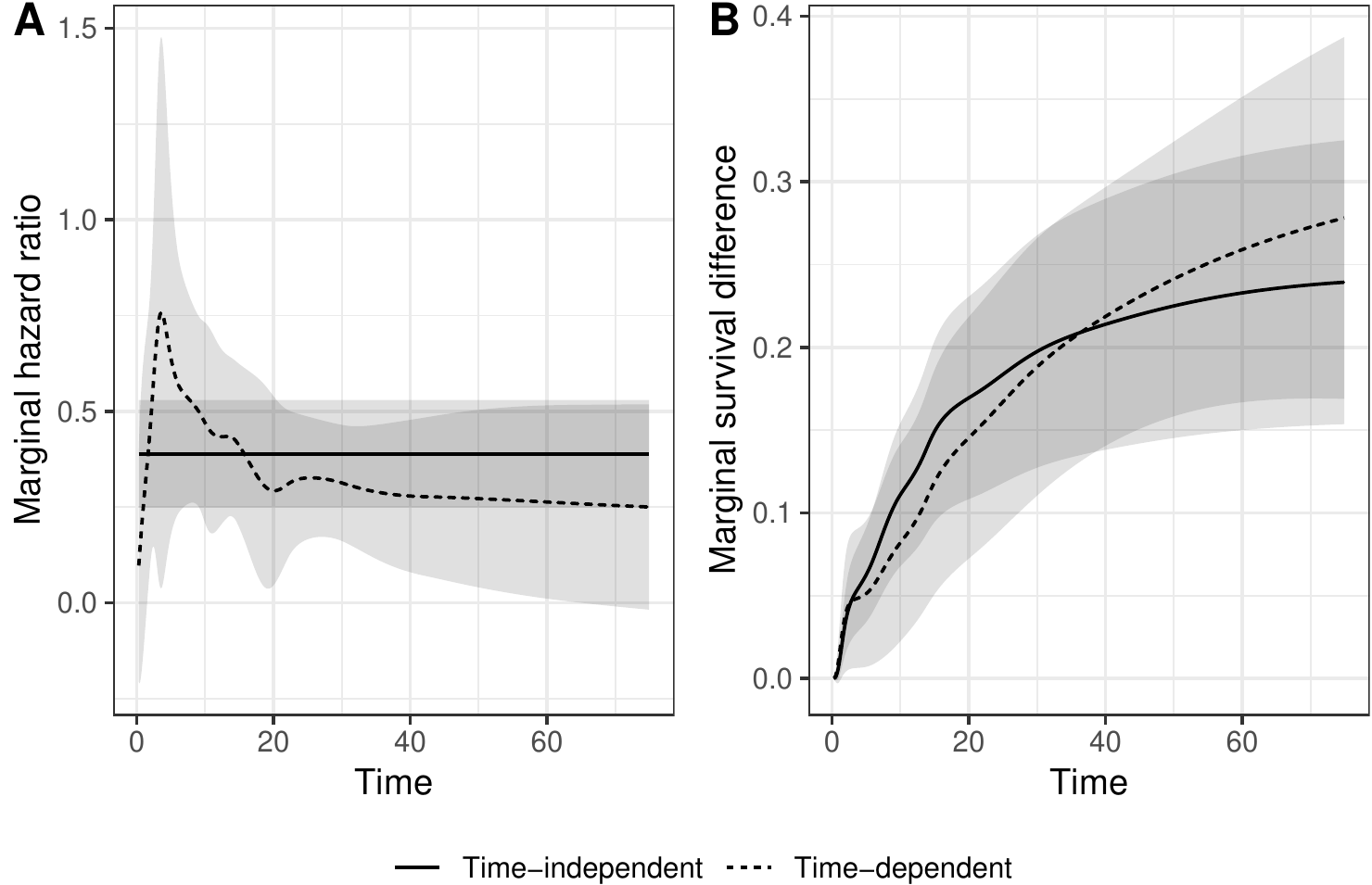}

}

\caption{Marginal hazard ratio (A) and marginal survival difference (B), comparing a flexible parametric model with 9 degrees of freedom and a log-Normal frailty with and without a time-dependent treatment effect. Application to diabetic retinopathy data.}\label{fig:drs-f-tvc}
\end{figure}

We include in flexible parametric model identified by the AIC as the
second best an interaction between treatment and the natural logarithm
of time, modelled using a natural spline: \[
\log H(t_{ij}) = s(\log(t_{ij}) | \gamma, k_0) + X_{ij} \beta + s(\log(t_{ij}) | \delta, l_0) X_{ij}  + \eta_i,
\] with the treatment variable \(X_{ij}\) is interacted with a spline
function of log-time with associated coefficient vector \(\delta\) and
knots vector \(l_0\). Flexible parametric models have been showed to be
insensitive to the number of knots utilised to model time-varying
effects, therefore we choose to use 3 degrees of freedom for simplicity
\citep{bower_2017}. The difference between the marginal hazard ratio
estimated using the model with a time-dependent treatment effect and the
model without is depicted in Figure \ref{fig:drs-f-tvc}, panel A. The
difference seems to be larger early on in the study, attenuating as time
goes by. As a comparison, we also include the marginal survival
difference between treated and untreated individuals, estimated assuming
both time-fixed and time-varying effects (Figure \ref{fig:drs-f-tvc},
panel B). Finally, we test whether the time-treatment interaction is
statistically significant using a likelihood ratio test. We obtain a
\(\chi^2\) test statistic of 2.87 and a p-value of 0.41: this suggest
that there is not enough evidence to support the presence of a
time-dependent treatment effect.

\hypertarget{application-bladder0}{%
\section{Application to bladder cancer
data}\label{application-bladder0}}

We also apply the same models from our simulation study to a dataset
constructed by pooling 7 trials on bladder cancer comparing chemotherapy
against no chemotherapy. The outcome of interest is time from
randomisation to cancer relapse, in years; patients still alive and
without recurrence were censored at the date of the last available
follow-up cystoscopy. We then compare estimates of relative and absolute
risk; specifically, we compare the hazard ratio and the 1-year LLE of
chemotherapy against no chemotherapy, respectively.

The dataset included 410 patients from 21 unique centres that
participated in EORTC trials. The median follow-up\sout{,} estimated
once again using the inverse Kaplan-Meier method \citep{schemper_1996}
and assuming robust standard errors was 3.52 years (confidence interval:
3.28 -- 3.88). The minimum and maximum observed follow-up times are 1
day and 10.15 years, respectively.

\begin{figure}

{\centering \includegraphics[width=0.8\textwidth]{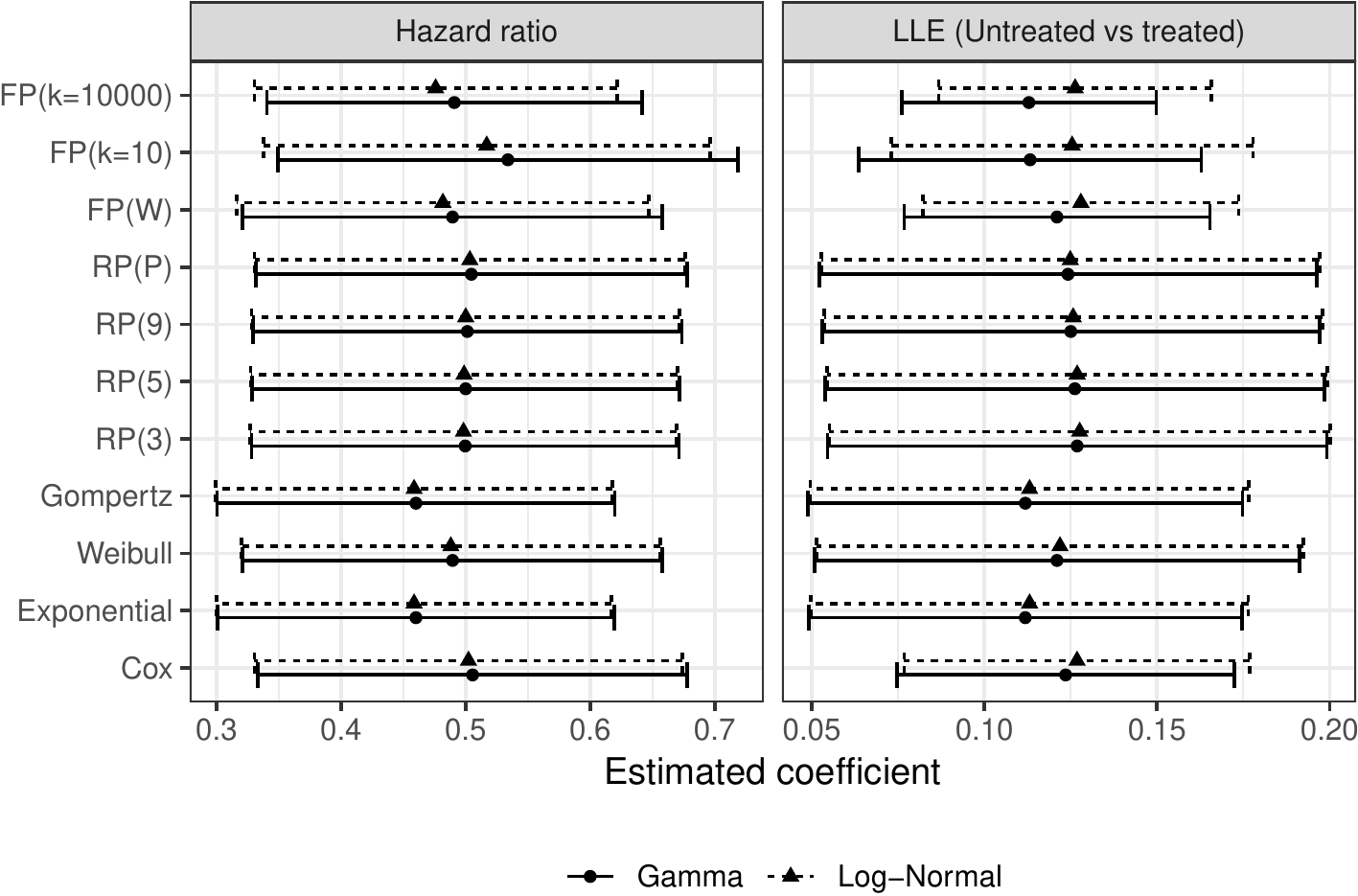}

}

\caption{Comparison of relative and absolute estimates of risk, application to bladder cancer data \label{fig:bladder0-forest-plot}}\label{fig:bladder0-forest-plots}
\end{figure}

\begin{table}[t]

\caption{\label{tab:bladder0-tab}Results: application of shared frailty models fitted to the bladder cancer data.}
\centering
\begin{tabular}{rccc}
\toprule
Baseline & Hazard ratio & Frailty variance & LLE (Untreated vs treated)\\
\midrule
\addlinespace[0.3em]
\multicolumn{4}{l}{\textbf{Gamma frailty:}}\\
\hspace{1em}Cox & 0.5056 (0.0880) & 0.0468 (0.0484) & 0.1235 (0.0249)\\
\hspace{1em}Exponential & 0.4601 (0.0812) & 0.0755 (0.0617) & 0.1118 (0.0320)\\
\hspace{1em}Weibull & 0.4895 (0.0860) & 0.0558 (0.0525) & 0.1210 (0.0358)\\
\hspace{1em}Gompertz & 0.4601 (0.0815) & 0.0755 (0.0620) & 0.1118 (0.0321)\\
\hspace{1em}RP(3) & 0.4997 (0.0876) & 0.0524 (0.0506) & 0.1269 (0.0369)\\
\hspace{1em}RP(5) & 0.5000 (0.0876) & 0.0522 (0.0500) & 0.1262 (0.0369)\\
\hspace{1em}RP(9) & 0.5014 (0.0878) & 0.0510 (0.0497) & 0.1250 (0.0367)\\
\hspace{1em}RP(P) & 0.5046 (0.0883) & 0.0478 (0.0488) & 0.1242 (0.0367)\\
\hspace{1em}FP(W) & 0.4895 (0.0860) & 0.0558 (0.0525) & 0.1210 (0.0226)\\
\hspace{1em}FP(k=10) & 0.5340 (0.0941) & 0.0416 (0.0457) & 0.1132 (0.0253)\\
\hspace{1em}FP(k=10000) & 0.4910 (0.0768) & 0.0610 (0.0526) & 0.1129 (0.0188)\\
\addlinespace[0.3em]
\multicolumn{4}{l}{\textbf{Log-normal frailty:}}\\
\hspace{1em}Cox & 0.5023 (0.0877) & 0.0640 (0.0593) & 0.1268 (0.0255)\\
\hspace{1em}Exponential & 0.4586 (0.0810) & 0.0891 (0.0744) & 0.1131 (0.0323)\\
\hspace{1em}Weibull & 0.4881 (0.0858) & 0.0634 (0.0608) & 0.1219 (0.0360)\\
\hspace{1em}Gompertz & 0.4586 (0.0813) & 0.0891 (0.0748) & 0.1131 (0.0324)\\
\hspace{1em}RP(3) & 0.4983 (0.0874) & 0.0591 (0.0580) & 0.1276 (0.0370)\\
\hspace{1em}RP(5) & 0.4988 (0.0874) & 0.0583 (0.0568) & 0.1269 (0.0370)\\
\hspace{1em}RP(9) & 0.5001 (0.0877) & 0.0570 (0.0564) & 0.1257 (0.0368)\\
\hspace{1em}RP(P) & 0.5034 (0.0882) & 0.0534 (0.0553) & 0.1249 (0.0368)\\
\hspace{1em}FP(W) & 0.4818 (0.0844) & 0.0787 (0.0585) & 0.1279 (0.0233)\\
\hspace{1em}FP(k=10) & 0.5169 (0.0915) & 0.0826 (0.0473) & 0.1254 (0.0268)\\
\hspace{1em}FP(k=10000) & 0.4760 (0.0743) & 0.1046 (0.0635) & 0.1263 (0.0202)\\
\bottomrule
\end{tabular}
\end{table}

Estimates values from each model fitted to the bladder trial data are
presented in Table \ref{tab:bladder0-tab} and Figure
\ref{fig:bladder0-forest-plots}. The estimated treatment effect obtained
from the semiparametric models and the flexible parametric models are
practically overlapping, with a\sout{n} hazard ratio of approximately
\(0.50\) (range: 0.4983 to 0.5056). Conversely, the parametric models
returned different results: the estimated relative risk was \(2-4\%\)
lower (e.g.~hinting towards a greater benefit of chemotherapy). Finally,
the penalised model fitted using the \texttt{frailtypack} package
returned quite different results depending on the penalisation parameter
\(\kappa\), with hazard ratios ranging between 0.4760 and 0.5340
(approximately 6\% risk difference), highlighting the importance of
choosing an appropriate smoothing parameter. Analogously, the estimates
of 1-year LLE were pretty consistent with each other and the
semiparametric and flexible model estimates were the most similar. Among
parametric models, the model with a Weibull baseline hazard seemed the
most similar to semiparametric and flexible parametric models; the
exponential and Gompertz models differed more, while penalised models
from \texttt{frailtypack} returned estimates close to those of flexible
parametric models (especially when assuming a log-Normal distribution
for the frailty). Overall, the 1-year LLE comparing unexposed and
exposed individuals varied between 0.1118 and 0.1279: practically
speaking, on average, individuals not exposed to chemotherapy
experienced recurrence of bladder cancer 1-1.5 months earlier during the
first year of follow-up.

\begin{table}[t]

\caption{\label{tab:bladder0-tab-AIC-BIC}Results: comparison of AIC/BIC from shared frailty models, application to bladder cancer data. Best AIC/BIC values are in bold.}
\centering
\begin{tabular}{rcc}
\toprule
Baseline & AIC & BIC\\
\midrule
\addlinespace[0.3em]
\multicolumn{3}{l}{\textbf{Gamma frailty:}}\\
\hspace{1em}Cox & --- & \vphantom{1} ---\\
\hspace{1em}Exponential & 959.57 & 971.62\\
\hspace{1em}Weibull & 951.45 & 967.52\\
\hspace{1em}Gompertz & 961.57 & 977.63\\
\hspace{1em}RP(3) & 884.69 & 908.78\\
\hspace{1em}RP(5) & 864.67 & 896.80\\
\hspace{1em}RP(9) & 858.16 & 906.36\\
\hspace{1em}RP(P) & --- & \vphantom{1} ---\\
\hspace{1em}FP(W) & 951.45 & 967.52\\
\hspace{1em}FP(k=10) & --- & \vphantom{1} ---\\
\hspace{1em}FP(k=10000) & --- & \vphantom{1} ---\\
\addlinespace[0.3em]
\multicolumn{3}{l}{\textbf{Log-normal frailty:}}\\
\hspace{1em}Cox & --- & ---\\
\hspace{1em}Exponential & 959.36 & 971.41\\
\hspace{1em}Weibull & 951.35 & 967.42\\
\hspace{1em}Gompertz & 961.36 & 977.43\\
\hspace{1em}RP(3) & 884.59 & 908.69\\
\hspace{1em}RP(5) & 864.59 & \textbf{896.72}\\
\hspace{1em}RP(9) & \textbf{858.08} & 906.28\\
\hspace{1em}RP(P) & --- & ---\\
\hspace{1em}FP(W) & 951.66 & 967.72\\
\hspace{1em}FP(k=10) & --- & ---\\
\hspace{1em}FP(k=10000) & --- & ---\\
\bottomrule
\end{tabular}
\end{table}

\begin{figure}

{\centering \includegraphics[width=0.4\textwidth]{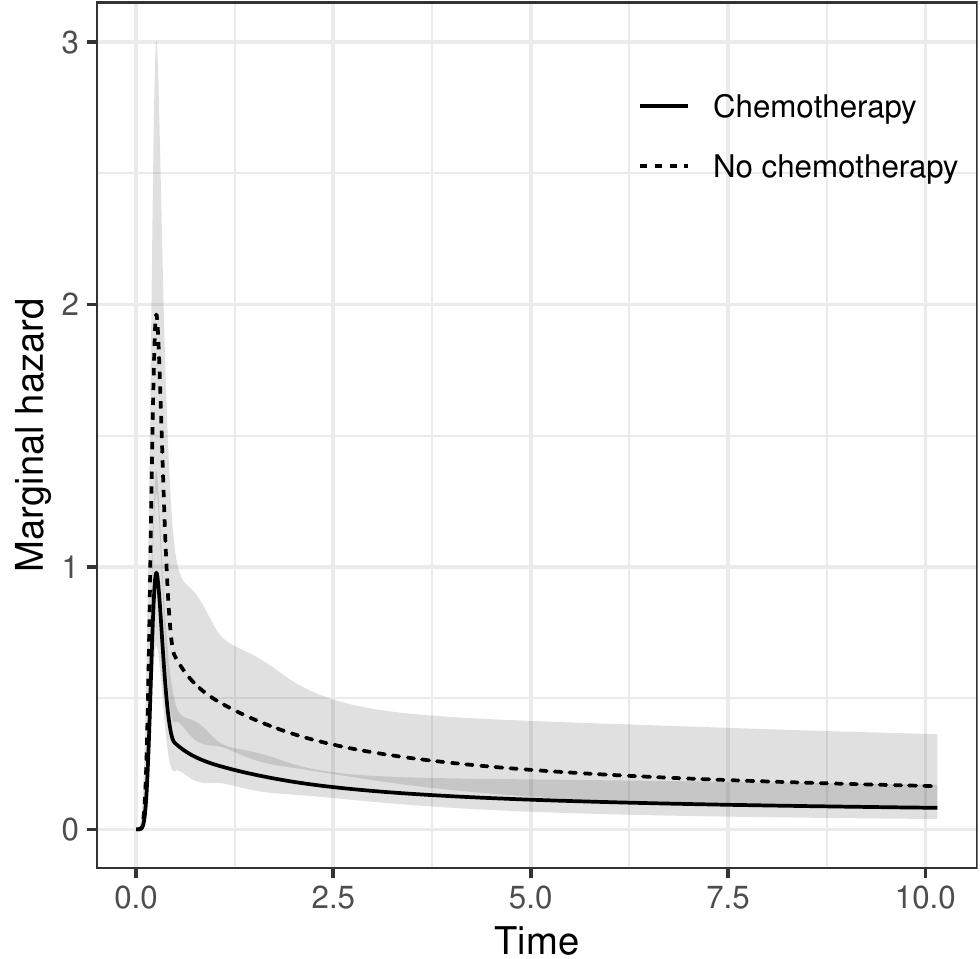}

}

\caption{Marginal hazard estimated from the flexible parametric model with 5 degrees of freedom and a log-Normal frailty. Application to bladder cancer data.}\label{fig:bladder0-est-hazard}
\end{figure}

Comparing model fit using AIC and BIC, the model with a flexible
baseline hazard provided the best fit according to both AIC and BIC.
Specifically, the model with 5 degrees of freedom for modelling the
baseline hazard and a log-Normal frailty was selected as the best model
by BIC. The penalised model with a flexible parametric baseline hazard
and a log-Normal frailty selected an effective number of degrees of
freedom of 11.75. Utilising this model to predict the marginal hazard,
we obtained Figure \ref{fig:bladder0-est-hazard}: first, individuals not
exposed to chemotherapy showed a higher hazard; second, we observed that
the hazard spiked early on and then decayed over time.

\begin{figure}

{\centering \includegraphics[width=0.8\textwidth]{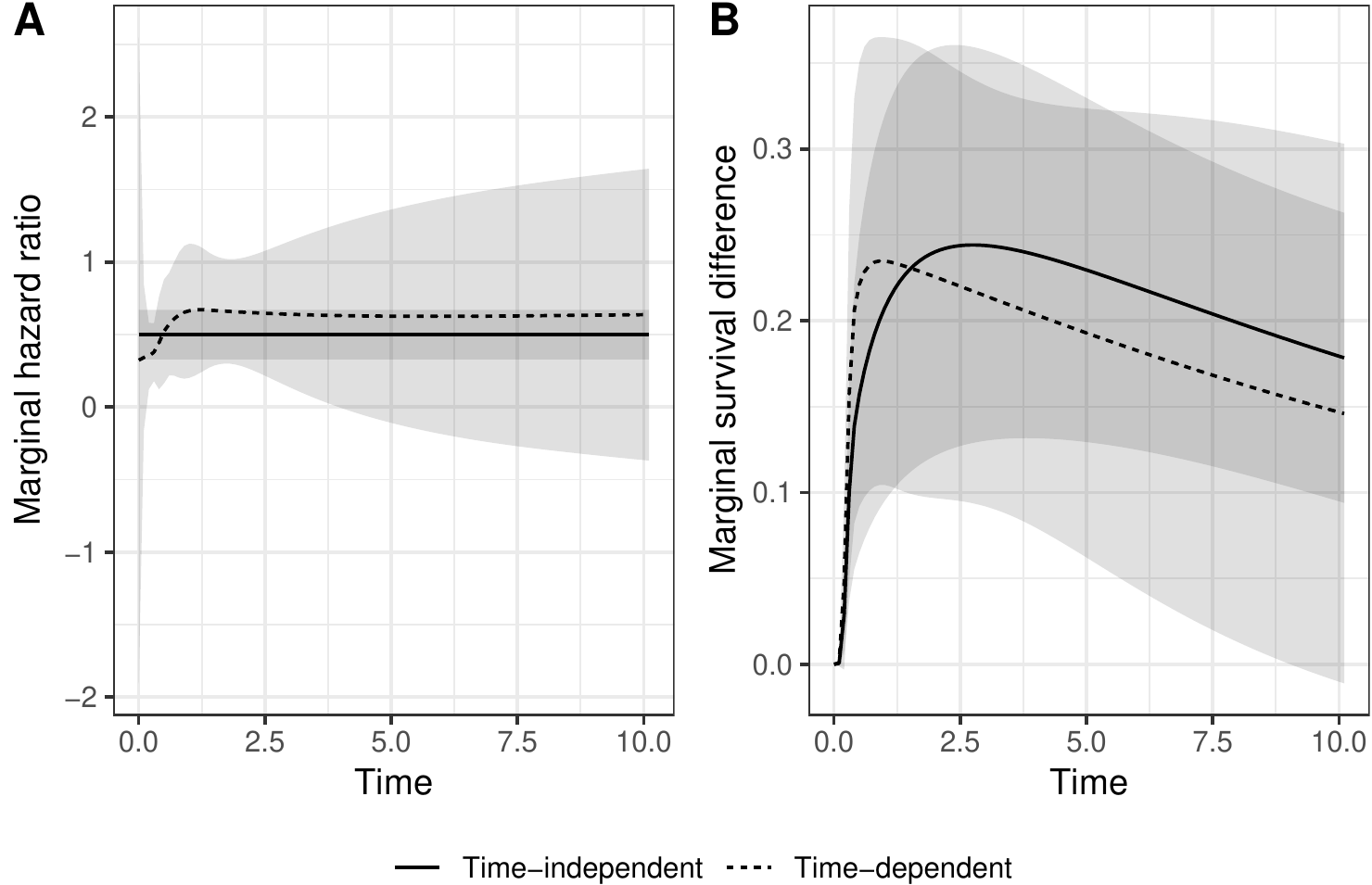}

}

\caption{Marginal hazard ratio (A) and marginal survival difference (B), comparing a flexible parametric model with 5 degrees of freedom and a log-Normal frailty with and without a time-dependent treatment effect. Application to bladder cancer data.}\label{fig:bladder0-f-tvc}
\end{figure}

Finally, analogously as the diabetic retinopathy applied example, it is
possible to test for a time-dependent effect of chemotherapy. We utilise
once again a natural spline with 3 degrees of freedom to model the
interaction between the logarithm of time and chemotherapy. The
estimated marginal hazard ratio and survival difference with and without
assuming a time-dependent treatment effect are presented in Figure
\ref{fig:bladder0-f-tvc}. The lines with estimated hazard ratios cross
at approximately 6 months of follow up: this showed a greater estimated
benefit of chemotherapy early on that lowers over time until it
converges to a risk reduction of approximately a \(30\%\) if we assume
the effect of chemotherapy to be time-dependent; conversely, the risk
reduction when assuming time-independent effects is approximately
\(50\%\). The same pattern can be observed with the marginal survival
difference. Finally, we can test the significance of the time-treatment
interaction using the likelihood ratio test: if we do so, we obtain a
\(\chi^s\) test statistic of 2.44 and a p-value of 0.49 -- evidence
against the presence of a time-dependent effect of chemotherapy.

\hypertarget{discussion}{%
\section{Discussion}\label{discussion}}

In observational studies and clinical trials with survival outcomes and
an intrinsic hierarchical structure, survival models with shared frailty
terms and/or random effects have moved from being a speciality rarely
used that requires ad hoc software to being mainstream methods that can
be utilised with any general-purpose statistical software such as
\texttt{R} and \texttt{Stata}. Compared to a marginal approach
(i.e.~accounting for clustering by using a robust estimator of the
variance-covariance matrix of the estimated coefficients), the frailty
approach allows focussing on inference within the clusters and
quantifying the amount of heterogeneity between clusters by directly
modelling it. Additionally, the frailty approach can be used to model
recurrent events data, assuming that the recurrent event times are
independent conditional on the covariates and random effects
\citep{hougaard_2000, amorim_2015}. Consequently, the adoption and use
of such models have been steadily increasing in all fields of
application: for instance, psychiatry \citep{psychiatry}, orthodontia
\citep{orthodontia}, diabetes \citep{diabetes}, healthcare research
\citep{healthcare}, leukaemia \citep{leukemia}. It is also increasingly
common to encounter survival data with some sort of hierarchical
structure; for instance, multi-centre clinical trials and individual
patient data meta-analysis, twin studies, paired organs studies, and
observational studies with geographical clustering. Another relevant
source of hierarchical data is electronic health records, where
individuals can be clustered within e.g.~primary care practice.

Glidden and Vittinghoff \citep{glidden_2004} showed the benefit of using
frailty models instead of models with fixed effects only or stratified
approaches in the setting of multi-centre clinical trials, which may
have driven adoption and use. Despite the increasing use of such
methods, however, there has been little research on the impact of
violating modelling assumptions - especially regarding the shape of the
baseline hazard. Much research has focussed on misspecification of the
frailty distribution, and the consensus is that relative risk estimates
are largely unaffected by it
\citep{pickles_1995, glidden_2004, lee_2008, liu_2017}. However, little
is known about the impact of misspecifying the frailty on measures of
absolute risk. With this simulation study, we aimed to shed further
light on the topic and ultimately provide additional guidance to applied
researchers.

We simulated clustered survival data under a variety of clinically
plausible scenarios, assuming different shapes for the baseline hazard
function and different distributions for the shared frailty. We varied
the amount of heterogeneity in the data we simulated, and we also varied
sample size, both in terms of number of clusters and number of
individuals per cluster. We then fitted a large variety of survival
models with shared frailty terms: assuming standard parametric
distributions for the baseline hazard, modelling the baseline hazard in
a flexible way via restricted cubic splines, and also leaving the
baseline hazard unspecified. Each model was fit assuming both a Gamma
and a log-Normal distribution for the frailty, arguably the most common
choices in literature: the Gamma frailty has convenient mathematical
features and it is analytically tractable, while the log-Normal frailty
has a direct interpretation as a random intercept in a multilevel
mixed-effects survival model. To the best of our knowledge, this is the
most extensive simulation study on the impact of misspecifying the
baseline hazard, the frailty distribution, or both in shared frailty
survival models: Rutherford \emph{et al}. only studied the robustness of
flexible parametric models and did not consider frailty terms, while
Pickles and Crouchley, Glidden and Vittinghoff, and Lee and Thompson
only studied misspecification of the random effects distribution
\citep{rutherford_2015, pickles_1995, glidden_2004, lee_2008}. Liu
\emph{et al}. focussed on generalised survival model \citep{liu_2017},
and Ha \emph{et al}. studied both misspecification of the baseline
hazard and the frailty distribution but included fewer models in their
comparison and only simulated a small amount of scenarios
\citep{ha_2003}.

The results of our extensive simulations confirm the robustness of
regression coefficients to misspecification of the frailty distribution,
irrespectively of sample size and amount of heterogeneity in the data.
However, our results also show the importance of modelling the baseline
hazard in a satisfactory way. For instance, as we showed in Section
\ref{results}, the bias induced by assuming a standard parametric
distribution with a true complex baseline hazard can be clinically
relevant. In practical terms, this means that by failing to model the
baseline hazard we could be largely overestimating (or underestimating)
the effect of interest. In the applied examples of Sections
\ref{application-drs} and \ref{application-bladder0}, the estimates of
relative risk differ by 3-6\% between the models - a difference that can
be significant in clinical terms. We showed that absolute measures of
risk such as the loss in expectation of life are affected by
misspecification of both the baseline hazard and the frailty
distribution: assuming a baseline hazard that is too simple or
misspecifying the frailty distribution yields biased estimates and
larger mean squared errors compared to well-specified models. In our
applied examples, the difference in estimated LLE between models varied
between 1 month (over 5 years) for the diabetic retinopathy example and
1-1.5 months (over 1 year) for the bladder cancer example. This
highlights once again the necessity of using models that are flexible
enough and the importance of assessing model fit regarding the
distribution of the frailty by using information criteria such as the
AIC and the BIC if no previous biological knowledge is available. In
addition to that, we once again encourage the maintainers of software
packages that can be used to fit shared frailty models to implement
additional information criteria adjusted to account for the presence of
the random effects, such as the cAIC \citep{ha_2007, donohue_2011}.
Comparing semiparametric Cox models and flexible parametric models, they
produce unbiased relative risk estimates under any of the scenarios
explored with this simulation study. However, the necessity of
estimating the baseline hazard (e.g.~by using the Breslow estimator)
heavily affects the usage of semiparametric models when absolute risk
measures are of interest. The Cox model is, de facto, the standard model
fitted by applied researchers when dealing with time to event data;
despite that, Sir David Cox himself argued in favour of parametric
models \citep{reid_1994}, especially when interested in predicting the
outcome for a given individual; parametric models are indeed known to
have desirable features in terms of prediction, extrapolation,
quantification of absolute risk measures. Flexible parametric models
represent an attractive alternative to semiparametric and fully
parametric survival models: they retain both the robustness to
misspecification of the baseline hazard and the appealing advantages of
parametric models for prediction, extrapolation, quantification. Since
their introduction by Royston and Parmar \citep{royston_2002} in 2002,
flexible parametric models have entered the statistical mainstream and
have been extended to accommodate (among other) relative survival
\citep{nelson_2007}, random effects \citep{crowther_2014, liu_2017}, and
generalised link functions \citep{liu_2018}. The advantage given to
flexible parametric models versus semiparametric models by modelling the
baseline hazard is particularly noteworthy: this allows translating
relative risk measures on the absolute scale, aiding interpretation.
With the two applications in Sections \ref{application-drs} and
\ref{application-bladder0} we illustrate in practice the importance of
choosing the right model, and how results are affected when that does
not happen - with clinically relevant differences in risk estimates.
Flexible parametric models produced consistent estimates, irrespectively
of the number of degrees of freedom used to model the baseline hazard;
this is consistent with previous results
\citep{rutherford_2015, bower_2017} and it is a key feature of this
class of models.

We mentioned in Section \ref{sim-dgms} that we simulated and explored 90
distinct data-generating mechanisms: this wide variety is one of the
advantages of this simulation study. Other advantages are: we included
the most common frailty distributions (Gamma and log-Normal), and we
simulated survival data under many different and clinically plausible
baseline hazards. For instance, should we only simulate from a Weibull
model, we would be assuming a baseline hazard that increases or
decreased monotonically. While such an assumption could be reasonable in
some settings, sometimes fully parametric distributions are just not
flexible enough to capture complex baseline hazards with turning points
that are often observed in clinical datasets
\citep{rutherford_2015, crowther_2013b}. This simulation study has also
some limitations. First, we only simulated clusters of equal size and we
did not include all the frailty distributions that have been proposed in
the literature. Second, we simulated only right censored survival data;
settings with delayed entry or interval censoring require further
investigation. Third, all methods use maximum likelihood which returns
negatively biased estimates of the variance components; such bias
decreases as the number of clusters increases. The restricted maximum
likelihood method could be used with a small number of clusters to
obtain unbiased estimates of the variance components
\citep{searle_2008}. However, but the comparison between maximum
likelihood and restricted maximum likelihood is outside of the scope of
this manuscript. Fourth, we designed and analysed this simulation study
using a fully factorial design; even though we simulated a large number
of scenarios, incomplete designs and meta-modelling could be implemented
to further increase the external validity and the ability to generalise
our results, as described elsewhere \citep{skrondal_2000}. Finally, we
heavily rely on the performances and R implementation of the models we
fit and compare; Hirsch and Wienke \citep{hirsch_2012} compared several
implementations of the semiparametric Cox model with frailty terms and
found \texttt{coxme} (our R package of choice for a semiparametric
log-Normal frailty model) to be among the most robust. Regardless, all
the packages we chose are well established and utilised in practice, and
we mimicked applied research by applying these methods as they are
intended to be used, i.e.~without modifying convergence criteria and/or
starting values of the estimation procedure.

\hypertarget{acknowledgements}{%
\section{Acknowledgements}\label{acknowledgements}}

MSC is supported by the Swedish eScience Research Centre. KRA is
partially supported by the UK National Institute for Health Research
(NIHR) as a Senior Investigator Emeritus (NF-SI-0512-10159). MJC is
partially funded by a Medical Research Council (MRC) New Investigator
Research Grant (MR/P015433/1).

\bibliography{bibl}

\end{document}